\documentclass{aa}
\usepackage{graphics}
\input{psfig}
\begin{document}

\thesaurus{02 (02.03.3; 02.04.2) 08 (08.02.6; 08.05.3; 08.06.3; 08.09.2)}

\title{ Calibrations of $\alpha$\,Cen\,A \& B}

\author{P. Morel\inst{1}, J. Provost\inst{1}, Y. Lebreton\inst{2},
F. Th\'evenin\inst{1} and G. Berthomieu\inst{1}}

\institute{
D\'epartement Cassini, UMR CNRS 6529, Observatoire de la C\^ote 
d'Azur, BP 4229, 06304 Nice CEDEX 4, France
\and 
DASGAL, UMR CNRS 8633, Observatoire de Paris-Meudon, 92195 Meudon
Principal CEDEX, France.}

\offprints{P. Morel}
\mail{Pierre.Morel@obs-nice.fr}

\date{Received date / Accepted date}

\maketitle

\begin{abstract}
Detailed evolutionary models of the visual binary
 $\alpha$\,Centauri, including pre main-sequence evolution,
 have been performed using the masses recently determined by Pourbaix et
 al.~(\cite{pnn99}). Models
 have been constructed using the CEFF equation of
 state, OPAL opacities, NACRE thermonuclear reaction rates and
 microscopic diffusion.
 A $\chi^2$-minimization is performed to derive the most reliable set of
 modeling parameters
 $\wp=\{t_{\rm\alpha\,Cen},Y_{\rm i},[{\rm\frac{Fe}H]_i},\alpha_{\rm A},\alpha_{\rm B}\}$,
 where  $t_{\rm\alpha\,Cen}$ is the
 age of the system, $Y_{\rm i}$ the initial helium content,
 $[{\rm\frac{Fe}H]_i}$ the initial metallicity and, 
 $\alpha_{\rm A}$ and $\alpha_{\rm B}$ the convection parameters
 of the two components. Using the basic B\"ohm-Vitense~(\cite{b58}) mixing-length
 theory of convection,
 we derive $\wp_{\rm BV}=\{\rm 2710\,Myr, 0.284,0.257,1.53,1.57\}$.
We obtain a noticeably smaller age than estimated previously, in 
agreement with Pourbaix et al.~(\cite{pnn99}), mainly because of the
larger masses. If convective core
 overshoot is considered we get $\wp_{\rm ov}=\{\rm 3530\,Myr, 0.279,0.264,1.64,1.66\}$. The use
 of Canuto \& Mazitelli~(\cite{cm91},~\cite{cm92}) convection theory leads to the set  
 $\wp_{\rm CM}=\{\rm 4086\,Myr, 0.271,0.264,0.964,0.986\}$. Using
 the observational constraints adopted by Guenther \& Demarque~(\cite{gd00}),
 and the basic mixing-length theory,  
  we obtain $\wp_{\rm GD}=\{\rm 5640\,Myr, 0.300,0.296,1.86,1.97\}$ and 
surface lithium depletions close to their observed values.

A seismological analysis of our calibrated models has been performed.
The determination of large and small spacings between the
frequencies of acoustic oscillations from seismic observations 
would help to discriminate between the models of $\alpha$\,Cen 
computed with different masses and to confirm or rules out the new determination 
of masses.

\end{abstract}

\section{Introduction}\label{sec:int} 
As the Sun's nearest stellar neighbors, the two members of the visual binary
$\alpha$\,Centauri\,A \& B (G2V + K1V)
provide the most accurate potentiality of testing stellar physics
in conditions slightly different from the solar ones and then deserve
undivided attention for internal structure modeling and
oscillation frequencies calculations.
By coincidence, the masses and the spectral types of
components A/B (HD 128620/1, IDS 14328-6025\,A/B, Hipparcos 71\,683/1)
bracket those of the Sun. 
The high apparent brightness and the large parallax
imply that surface abundances
and astrometric parameters are known better than for any star (except the
Sun. The basic intent of this paper is to model
$\alpha$\,Cen\,A \& B using updated physics. 
This allows to predict of the $p$-mode oscillation
frequencies, which will be useful to exploit future asteroseismological
observations as expected, for instance, from the project
Concordiastro at the South Pole (Fossat et al.~\cite{fgv00}).
Also, as ``solar like stars'', the two components are primary
targets for the MONS (Kjeldsen et
al.~\cite{kbcd99}) spatial mission and their
oscillations are expected to be well
separated in the frequency spectrum.

Based on the reasonable hypothesis of a common origin for both components,
i.e. same initial chemical composition and age,
the calibration of a binary system consists in determining a consistent
evolutionary history for the double star, given (1) the positions
of the two components in a H-R diagram, (2) the stellar masses and,
(3) the present day surface chemical composition. The goal is to
compute evolutionary models that reproduce the observations.
This procedure yields estimates for the age $t_\star$, the initial helium
mass fraction $Y_{\rm i}$ and initial metallicity $\rm [\frac{Fe}H]_{\rm i}$
(logarithm of the number abundances of iron to hydrogen relative to the
solar value),
which are fundamental quantities for our understanding of the galactic chemical
evolution. We also derive values of the ``mixing-length parameter'' or ``convection
parameter'' $\alpha$, ratio of the mixing-length to the pressure
scale height.
Once the initial masses and the physics are fixed, the modeling
of the two components A \& B of a binary system
requires a set $\wp$ of five so-called modeling parameters:
\[\wp=\left\{t_\star, Y_{\rm i}, \rm [\frac{Fe}H]_i,
\alpha_{\rm A}, \alpha_{\rm B}\right\}.\]
In most cases there are only four reliable
observables namely, the effective temperatures
$T_{\rm eff\,A},\,T_{\rm eff\,B}$ and the luminosities
$L_{\rm A},\,L_{\rm B}$ or the gravity
$\log g_{\rm A},\,\log g_{\rm B}$ of each component.
Therefore, one of the unknowns has to be fixed. Very often the mixing-length
parameters are assumed to be the same for both 
components, even if the mass ratio differs
significantly from unity. 
Once detailed spectroscopic analyses have been performed on the
system, the precise present day surface metallicities of stars
come as additional observational constraints.

In this paper we attempt to reproduce the observed metallicities
and, if possible, the lithium depletion by means of models
including  microscopic diffusion. We calibrate the binary system using
both  B\"ohm-Vitense's~(\cite{b58}, hereafter $\rm MLT_{BV}$)
and Canuto \& Mazitelli~(\cite{cm91},~\cite{cm92}, hereafter $\rm MLT_{CM}$)
 mixing-length convection theories.

The paper is divided as follows: in Sect.~\ref{sec:ptw}, we recall the
main results obtained in previous theoretical works and, in
Sect.~\ref{sec:mlt},
we emphasize the difficulties related to the choice of 
mixing-length parameters. 
In Sect.~\ref{sec:tp}, we discuss the modeling of the transport
processes acting beneath the convection zone.
The observational material, relevant to the evolutionary status of $\alpha$\,Cen
and available in the literature, is collected in Sect.~\ref{sec:acen}.
The method of calibration is described in Sect.~\ref{sec:chi2}.
In Sect.~\ref{sec:comp}, we present the stellar modeling procedure.
In Sect.~\ref{sec:res}, we give the results with emphasis on the seismological
analysis. We summarize our results and conclude in Sect.~\ref{sec:dis}.

\begin{table*}
\caption[]{
Modeling parameters and main characteristics of the oscillation spectrum
of $\alpha$\,Cen\,A \& B derived in this study
 and taken from the literature. The symbols have
 their usual meaning (see text). The references are:
(1) this paper, models $\rm A_{BV}$ \& $\rm B_{BV}$,
(2) this paper, models $\rm A_{ov}$ \& $\rm B_{ov}$,
(3) this paper, models $\rm A_{CM}$ \& $\rm B_{CM}$,
(4) this paper, models $\rm A_{GD}$ \& $\rm B_{GD}$,
(5) Flannery \& Ayres~(\cite{fa78}),
(6) Demarque et al.~(\cite{dga86}),
(7) Noels et al.~(\cite{ngmnbl91}),
(8) Edmonds et al.~(\cite{ecdgp92}),
(9) Neuforge~(\cite{n93}),
(10) Lydon et al.~(\cite{lfs93}),
(11) Pourbaix et al.~(\cite{pnn99}),
(12) Guenther \& Demarque~(\cite{gd00}).
}\label{tab:glob}
\begin{tabular}{llllllllllllll}  \hline \\
$\alpha_{\rm A}$         &$\alpha_{\rm B}$         &$Y_{\rm i}$              &$\rm[\frac{Fe}H]_i$      &$(\frac ZX)_{\rm i}$&$t_{\rm\alpha\,Cen}$           &$\rm\Delta\nu_0^A$&$\rm\overline{\delta\nu}_{0 2}^A$&$\rm\Delta\nu_0^B$&$\rm\overline{\delta\nu}_{0 2}^B$\\
&&&&&Myr&$\mu$\,Hz&$\mu$\,Hz&$\mu$\,Hz&$\mu$\,Hz\\ \\
\hline \\
$1.53_{-0.06}^{+0.06}$   &$1.57_{-0.11}^{+0.03}$   &$0.284_{-0.002}^{+0.002}$&$0.257_{-0.002}^{+0.002}$&0.0443         &$2710_{-690}^{+550}$&108            &8.9                & 154           & 12.5          &(1)\\ \\
$1.64_{-0.04}^{+0.06}$   &$1.66_{-0.01}^{+0.04}$   &$0.279_{-0.002}^{+0.002}$&$0.264_{-0.002}^{+0.002}$&0.0450         &$3530_{-690}^{+550}$&107            &9.1                & 154           & 12.5          &(2)\\ \\
$0.96_{-0.02}^{+0.02}$   &$0.99_{-0.09}^{+0.06}$   &$0.271_{-0.004}^{+0.002}$&$0.264_{-0.004}^{+0.004}$&0.0450         &$4090_{-690}^{+410}$&108            &7.5                & 157           & 11.7          &(3)\\ \\
$1.86_{-0.06}^{+0.09}$   &$1.97_{-0.15}^{+0.13}$   &$0.300_{-0.002}^{+0.002}$&$0.296_{-0.002}^{+0.003}$&0.0480         &$5640_{-210}^{+210}$&102            &4.5                & 161           & 10.6          &(4)\\
\\  \hline \\
1.33                     &1.33                     &0.246                    &                         &0.04           &$6000\pm1000$       &               &                   &               &               &(5)\\ 
1.37                     &1.37                     &0.236                    &                         &0.02-0.04      &$3500-5500$         &$118$          &                   &               &               &(6)\\ 
1.6                      &1.6                      &0.32                     &                         &0.04           &$5000\pm500$        &               &                   &               &               &(7)\\ 
1.15                     &1.25                     &$0.300\pm0.005$          &                         &$0.026\pm0.003$&$4600\pm400$        &108            & 6.2               & 179           & 12.6          &(8)\\
2.10                     &$2.10$                   &0.321                    &                         &0.038          &4840                &               &                   &               &               &(9)\\
$1.70\pm0.1$             &$2.1\pm0.1$              &$0.298\pm0.006$          &                         &$0.031\pm0.002$&$5600\pm500$        &               &                   &               &               &(10)\\
$1.86\pm0.8$             &$2.1\pm0.8$              &$0.284\pm0.058$          &                         &$0.030$        &2700                &$107$          &$9$                &               &               &(11)\\
$2.33$                   &$2.54$                   &$0.280$                  &                         &0.034          &$7600\pm800$        &$101$          & $4.6$             & $173$         & $15$          &(12)\\
\\ \hline
\end{tabular}
\end{table*}

\section{Previous theoretical works}\label{sec:ptw}
On the basis of rotational velocity measurements,
observed \element[][]{Li} abundances and \element[][]{CaII} emission
intensity, Boesgaard \& Hagen~(\cite{bh74}) attributed to the system an age of
$t_{\rm\alpha\,Cen}=3.6$\,Gyr.
Then, several groups calculated stellar evolution models to
calibrate the system and draw
information on the two components as well as on the
physics governing their structure. 
Table~\ref{tab:glob} gives the values of the calibration
parameters derived in all those studies which we now briefly
summarize.

Initially, Flannery \& Ayres~(\cite{fa78}) and 
Demarque et al.~(\cite{dga86}) could only use the luminosities of 
$\alpha$\,Cen\,A \& B as observational constraints.
Flannery \& Ayres
models support the fact that the system is metal rich
with respect to the Sun with $Z_{\rm \alpha\,Cen}\sim 2Z_\odot$.
Demarque et al.~(\cite{dga86}) derived the age of the system as a function of
metallicity; they also computed the $p$-mode oscillation
spectrum of $\alpha$\,Cen\,A.
Noels et al.~(\cite{ngmnbl91}) introduced a general procedure for fitting
models to the binary system. 
 They derived the age, the helium content
and the metallicity of the system and the value of the $\rm MLT_{BV}$
parameter assuming that
it is the same for the two stars. Neuforge~(\cite{n93}) revisited
that work using OPAL opacities (Iglesias et al.~\cite{irw92})
complemented by her own low-temperature opacities.
 Fernandes \& Neuforge~(\cite{fn95}) showed that the
mixing-length parameter $\alpha$ obtained through calibration is different
for the two stars and that their values become very similar if
the mass fraction of heavy elements increases above $Z\sim0.035$.
They also performed model calibrations with the $\rm MLT_{CM}$
convection treatment with a 
mixing-length equal to the distance to the top of the convective
envelope, thus avoiding the calibration of a convection
parameter.
In parallel, Edmonds et al.~(\cite{ecdgp92}) were the first to add
the observed metallicity as a constraint
(releasing in turn the hypothesis that $\alpha$ is unique)
and to include the effect of microscopic diffusion. 
More recently, Pourbaix et al.~(\cite{pnn99}) revisited the calibration of 
the $\alpha$\,Cen system. They calculated a new visual orbit
on the basis of available separations, position angles and
precise radial velocities
measurements and derive new consistent values of the orbital
parallax, sum of masses,
mass ratio and individual masses. The main result is that the
masses of the components are 5\% higher and that the
helium abundance and age are significantly smaller than previous
estimates. Guenther \& Demarque~(\cite{gd00}) performed several
calibrations of the system using different values of
the parallax including the Hipparcos value and models
calculated with updated physics including helium and heavy
elements diffusion. They
estimated the uncertainties on the calibrated parameters
resulting from the error bars on mass, luminosity, effective
temperature and chemical composition.

The differences between these calibrations reflect the great
improvements in the description of the stellar micro-physics 
achieved during the last two decades and the
progress of the analysis of the observational data. However,
uncertainties remain on the transport processes (e.g.
convection, microscopic and turbulent diffusion).
Apparently none of the previous calibrations has attempted to
reproduce the observed 
surface metallicities (except Guenther \& Demarque~\cite{gd00}) and lithium
abundances which are fundamental data for the
understanding of the transport processes beneath the convection
zone of solar-like stars.

As  asteroseismology will in a near future strongly constrain the
stellar models, some of the above described theoretical works give
the main characteristics of the oscillation spectrum of the components
($\Delta\nu_0$ and
$\overline{\delta\nu}_{0 2}$ -- see definitions Sect.~\ref{sec:pmode}).
The ``mean'' large
separation $\Delta\nu_0$ between the frequencies of modes of a
given degree and of consecutive radial order
depends on the stellar radius and mass. The ``mean'' small
separation $\delta\nu_{0 2}$  between the frequencies of modes
with  degree $\ell = 0$ and 2 and consecutive radial order, measured by 
$\overline{\delta\nu}_{0 2}$, is sensitive to the structure
of the stellar core. Table~\ref{tab:glob}
gives estimates of these quantities from the literature and from this work.

\section{The puzzle of the mixing-length parameter}\label{sec:mlt}
The calibration of a solar model provides the value of the solar
mixing-length parameter $\alpha_\odot$ which in turn is currently used to model
other stars with the same input physics. The value of
 $\alpha_\odot$ changed in time because of
successive updates of the input physics, in particular the
low-temperatures opacities and model atmospheres.
Apart from the difficulty of evaluating
$\alpha_\odot$, the question arises whether stars of different masses,
initial chemical composition and evolutionary status can be
modeled with a unique $\alpha$.
 Because models are also used to calculate isochrones, it is
important to try to clarify the situation: 
if $\alpha$ is proved to vary significantly with mass, metallicity and with
evolution, the isochrones might have quite different shapes which
could modify the age estimates.
 On the other hand, models of various masses,
all computed with $\alpha_\odot$, can reproduce quite well the slope
of the main-sequence of the Hyades cluster (Perryman et al.~\cite{pblgt98})
and of field stars (Lebreton et al.~\cite{lpcbf99}) observed by
Hipparcos, indicating that $\alpha$ does not vary much
for masses close to the solar mass.

For binaries with well-known properties as $\alpha$\,Cen,
 the question of the universality of the  
mixing-length parameter has  been examined many times.
Table~\ref{tab:glob} shows that
among the attempts of calibrating $\alpha$\,Cen\,A \&
B in luminosities and radii using models based on the $\rm MLT_{BV}$, two yield
values of $\alpha$ different for each component and different from 
$\alpha_\odot$ (Lydon et al.~\cite{lfs93}, Pourbaix et al.~\cite{pnn99})
while other calibrations suggest similar values for the two stars
(Edmonds et al.~\cite{ecdgp92}, Fernandes \& Neuforge~\cite{fn95}).
Also, Fernandes et al.~(\cite{flbm98}) calibrated three other binary
systems and
the Sun with the same input physics and concluded that
$\alpha$ is almost constant for $\rm[\frac{Fe}H]$ in the range
$\rm [\frac{Fe}H]_\odot
\pm\ 0.3$\,dex and masses in the range  $0.6-1.3\,M_\odot$ 
while Morel et al.~(\cite{mmpb00}) found a small difference of $\alpha$ 
($\approx 0.2$) in the two components of the $\iota$\,Peg
system.

As pointed out by many authors (e.g. Lydon et al.~\cite{lfs93},
Andersen~\cite{a91}) accurate masses and 
effective temperatures are required to draw firm 
conclusions on
whether the $\rm MLT_{BV}$ parameters are different or not. This is also
illustrated by the large error bars on $\alpha$ ($\approx 0.8$) quoted by
Pourbaix et al.~(\cite{pnn99}) which include all possible
observable sources of errors including the errors in the masses.
For a detailed estimate of the error budget see
e.g. Guenther \& Demarque~(\cite{gd00}).

Some 2-D and 3-D numerical simulations of convection have been performed
and translated into effective mixing length parameter
(Abbett et al.~\cite{abdgrs97}, Ludwig et
al.~\cite{lfs99}, Freytag et al.~\cite{fls99}). 
They suggest that, for stars with effective temperature, gravity and metallicity
 close to solar the mixing-length parameter remains almost constant.
Similar results are obtained by Lydon et al.~(\cite{lfs93}) who compared models of
$\alpha$\,Cen\,A \& B based on the results of numerical simulations of
convection and found that a  single value of
  $\alpha$ may be applicable to main sequence solar type stars.
In the $\rm MLT_{CM}$ convection theory,
Canuto \& Mazitelli~(\cite{cm91}, \cite{cm92})
used two modern theories of turbulence to establish
a formula for the turbulent convective flux which  replaces 
the $\rm MLT_{BV}$ expression (because the full spectrum of convective eddies
 is taken into account, the convective efficiency is 
magnified by about one order of magnitude). The $\rm MLT_{CM}$ theory 
also makes use of a mixing length which is equal, either to the distance
 towards the outermost limit of the convection zone,
or to the fraction ($\alpha$) of the pressure scale height.
The solar calibration with the $\rm MLT_{CM}$ treatment yields 
an $\alpha_\odot$ value of the order of unity.
 It is to be noted that for the Sun, models calibrated with $\rm MLT_{CM}$
 have frequencies which are closer 
to the observations than those calibrated with $\rm MLT_{BV}$
(Christensen-Dalsgaard et al.~\cite{jcd96}).

\section{Diffusion and transport processes}\label{sec:tp} 
For the Sun, it is now well established that microscopic 
diffusion by gravitational 
settling has a measurable impact on models thus on their pulsation spectrum.
As time goes on, all chemical species sink in the gravitational well,
except the lightest one, \element[][1]{H}. At the surface, the amount 
of hydrogen is enhanced while helium and heavy 
elements are depleted. This implies a decrease of the surface metallicity 
$\rm[\frac{Fe}H]_{\rm s}$ with respect to time.
Since the time-scale of abundance variations in the convection zone is roughly 
proportional to the square root of the mass of the convection zone (Michaud 
\cite{m77}), the microscopic diffusion time-scale decreases with
increasing stellar mass implying that diffusion 
is more efficient in massive stars. As a consequence, a 
smaller surface metallicity is expected for 
$\alpha$\,Cen\,A than for $\alpha$\,Cen\,B. This is an additional 
constraint for the models. 

As lithium burns at temperatures close to 
3\,MK, the observed surface abundance $\rm [Li]_s\ (H\equiv12)$
is often used as a probe for theories of transport 
processes at work beneath the external convective zone. 
The microscopic diffusion alone is 
not efficient enough to explain the lithium depletion 
observed in the Sun and stars; thus,
there is a need for an unknown physical process,
e.g. a turbulent mixing, acting in the radiative zone just 
beneath the outer convection zone -- e.g. Schatzman~(\cite{s96}),
Mont\`alban \& Schatzman~(\cite{ms00}) for a review.
According to the present state of the art, 
either the shear resulting from the different rotational status between the 
convection zone and the radiative interior (e.g. Zahn~\cite{z92}), or 
internal waves (e.g. Mont\`alban \& Schatzman~\cite{ms00} ),
are believed to be responsible for that turbulence. 
For stars more massive than the Sun, turbulent mixing is required, in some 
(hypothetical) mixed stellar mass fraction, to connect
and to extend somehow the mixing 
of the external convection zones. This 
avoids the complete segregation of helium and heavy elements from the surface 
and can explain the AmFm phenomenon 
(e.g. Richer et al.~\cite{rmt00}). 
Despite numerous attempts, there is today no fully satisfactory 
prescription for turbulent diffusion able to account for the observed 
solar lithium depletion; for other stars the situation is even 
worse\,! Further information on the mixing processes at work below
the outer convection zones can be provided by simultaneous
observations of lithium and beryllium, this latter being depleted
deeper in the stars at temperatures slightly larger than for lithium.

\section{Observations of the visual binary $\alpha$\,Cen}\label{sec:acen}
The calibration of a binary system is based on the adjustment
of the stellar modeling parameters, so that the model of each component
at time $t_{\rm\alpha\,Cen}$
reproduces the available observables within their error bars.
Among different possibilities, one has to choose the most suited
set of variables to be used in the calibration according
to the observational and theoretical material available.
As aforementioned, because of the assumption of a common origin of both
components,
five unknowns enter the modeling of their evolution.
There are six most currently available constraints: (1) the
effective temperatures for $\alpha$\,Cen\,A \& B, derived
from precise detailed spectroscopic analysis,
(2) either the spectroscopic gravities 
or the luminosities, these latter derived from the
photometric data and parallax, using bolometric corrections
(3) the surface metallicities. In the following, we discuss all the relevant
available observations and we choose the most appropriate and accurate
observables to be fitted in the calibration process.

\begin{table*}
\caption[]{ 
Astrometric properties of the $\alpha$\,Centauri A \& B binary system.
$P$ is the orbital period in yr, $a$ the semi major axis in arc second,
$\varpi$ the parallax in mas,
$K=M_{\rm B}/(M_{\rm A}+M_{\rm B})$ the fractional mass.
}\label{tab:dataa}
\begin{tabular}{llllllllllll} \hline \\
$P$           &$a$           &$\varpi$     &$K$           &$M_{\rm A}/M_\odot$&$M_{\rm B}/M_\odot$&References\\
\\  \hline \\
$79.92$       &$17.515$      &750          &0.453         &$1.09$                    &$0.90$                    &Heintz~(\cite{h58}),~(\cite{h82})\\
              &              &$749\pm5$    &$0.454\pm0.02$&$1.10$                    &$0.91$                    &Kamper \& Wesselink~(\cite{kw78})\\
              &              &$750.6\pm4.6$&              &                          &                          &Demarque et al.~(\cite{dga86})\\
              &              &$742.1\pm1.4$&              &                          &                          &ESA~(\cite{e97})\\
$79.90\pm0.01$&$17.59\pm0.03$&$737.0\pm2.6$&$0.45\pm0.1$  &$1.16\pm0.03$             &$0.97\pm0.03$             &Pourbaix et al.~(\cite{pnn99}) (orbital parallax)\\
              &              &$747.1\pm1.2$&              &                          &                          &S\"oderhjelm~(\cite{s00})\\
\\  \hline
\end{tabular}
\end{table*}

\subsection{Astrometric data}
For a visual binary, the parallax uncertainty is the 
largest source of error in the derivation of the sum of
masses via the third Kepler law (Couteau~\cite{c78}, 40, VI).
Unfortunately, $\alpha$\,Cen was ``difficult'' for Hipparcos with
the secondary at the edge of the sensitivity profile (S\"oderhjelm~\cite{s00}).
S\"oderhjelm improved the adjustment of the Hipparcos parallax
by taking into account the known orbital motion but, as expected,
the orbital curvature covered during the 3.25\,yr mission,
was not sufficient to derive a reliable mass ratio of the components.
Table~\ref{tab:dataa}
shows that the new parallax value differs from the
former by more than $2\sigma$. With respect to the original Hipparcos
value this new one is closer to the long focus photographic
determination of Kamper \& Wesselink~(\cite{kw78})
based on Heintz's~(\cite{h58}) orbital elements.

After numerous independent investigations during almost 100 years,
Pourbaix et al.~(\cite{pnn99}) derived an orbit based both
on visual astrometric and
precise radial velocity data, leading to consistent
determinations of orbital parameters, orbital parallax, sum of
masses, mass ratio and, thus, precise
mass values for $\alpha$\,Cen\,A \& B. Here, we prefer to
use of Pourbaix et al.~(\cite{pnn99}) mass values because 
S\"oderhjelm's~(\cite{s00}) ones are based on disparate
sources (i.e., Hipparcos parallax, Heintz's~(\cite{h58}) orbit 
and the mass ratio of Kamper \& Wesselink~(\cite{kw78})). 
We do not take into account mass errors in the calibrations.

The inclinations of the orbits of Heintz~(\cite{h58},~\cite{h82}) and 
Pourbaix et al.~(\cite{pnn99}) agree within less than one tenth of a degree,
 $i=79\degr29\pm0\degr01$, $\sin i\simeq1$ (the Sun lies close to the
orbital plane of $\alpha$\,Cen).
With the reasonable assumption that the orbital and rotational axis are
parallel, owing to the high inclination, $\alpha$\,Cen\,A \& B
are seen about equator on and the observed rotational
velocities are close to their equatorial values, $v\sin i\simeq v$.

\subsection{Photometric data}\label{sec:p}
Photometric data are available in the $V$, $B$ (Mermilliod et al.~\cite{mmh97}) 
and $K$ (Thomas et al.~\cite{thr73}) bands. But, as claimed by
Flannery \& Ayres~(\cite{fa78}) 
and Chmielewski et al.~(\cite{cfcb92}), the dynamical range of commonly
used photometers is limited, which does not allow precise photometric
measurements of bright stars. This makes the $\alpha$\,Cen photometric data rather
unsafe (Chmielewski et al.~\cite{cfcb92}) and we preferred to use the
spectroscopic data.  However, for comparison, we have used
these photometric data  to derive bolometric corrections and effective
temperatures of the two components from (1) the empirical calibrations
of Alonso et al.~(\cite{aam95}, \cite{aam96}) and (2) the theoretical calibrations of
Lejeune et al.~(\cite{lcb98}).  The $T_{\rm eff}(V-K,\rm [\frac{Fe}H])$ empirical
calibration of Alonso et al.~(\cite{aam96}) leads to
$T_{\rm eff,\,A}= 5738\pm40$\,K and $T_{\rm eff,\,B}=5165\pm70$\,K
while the theoretical (based on model
atmospheres) calibrations of Lejeune et al.~(\cite{lcb98}) 
yield $T_{\rm eff,\,A}\simeq5800$\,K and $T_{\rm eff,\,B}\simeq5250$\,K,
all well compatible with
spectroscopic temperatures. The $F_{\rm bol}(V-K,\rm
[\frac{Fe}H])$ empirical calibration of the bolometric flux
(Alonso et al.~\cite{aam95}) provides
 bolometric corrections ${\rm BC_A}(V)=-0.09$ and ${\rm BC_B}(V)=-0.23$
close to the values inferred from the theoretical calibrations of
Lejeune et al.~(\cite{lcb98}) ($-0.11$ and $-0.23$ respectively).
Let us point
out that these bolometric corrections are larger than those adopted by
Guenther \& Demarque~(\cite{gd00}).

\begin{table*}
\caption[]{
Spectroscopic data of the $\alpha$\,Cen system. The effective temperatures are in K,
the metallicities and the lithium abundance in dex.
For sake of brevity the uncertainties are not recalled.
}\label{tab:datas}
\begin{tabular}{lllllllllll} \hline \\
$\rm T_{eff\,A}$ & $\rm T_{eff\,B}$ & $\rm \log g_A$ & $\rm \log g_B$ & $\rm [\frac{Fe}H]_A$  
& $\rm [\frac{Fe}H]_B$ & $\rm [Li]_A$ & $\rm [Li]_B$ & References\\ 
\\ \hline \\
5793 & 5362 &      &      & 0.22 & 0.12 &&& French \& Powell~(\cite{fp71}) \\  
5727 & 5250 & 4.38 & 4.73 & 0.29 & 0.37 &&& England~(\cite{e80}) \\
5820 & 5350 &      &      &$-0.01$&$-0.05$&&& Bessell~(\cite{b81}) \\
     &      &      &      &      &      &$1.28$&&Soberblom \& Dravins~(\cite{sd84}) \\ 
5720 &      & 4.27 &      & 0.15 &      &&& Rabolo et al.~(\cite{rccfb86}) \\ 
5793 & 5305 & 4.40 & 4.65 & 0.20 & 0.20 &&& Smith et al.~(\cite{sef86}) \\ 
5793 & 5305 & 4.50 & 4.50 & 0.20 & 0.14 &&& Abia et al.~(\cite{arbc88}) \\  
5727 & 5250 & 4.42 & 4.65 & 0.20 & 0.26 &&& Edvardsson~(\cite{e88}) \\  
5727 &      & 4.27 &      & 0.10 &      &&& Furenlid \& Meylan~(\cite{fm90}) \\  
5800 & 5325 & 4.31 & 4.58 & 0.22 & 0.26 &1.3&$\leq 0.4$& Chmielewski et al.~(\cite{cfcb92}) \\  
5720 &      & 4.27 &      & 0.15 &      &&& Edvardsson et al.~(\cite{eaglnt93}) \\  
5830 & 5255 & 4.34 & 4.51 & 0.25 & 0.24 &&& Neuforge-Verheecke \& Magain~(\cite{nm97}) \\
     &      &      &      &      &      &$1.37$&&King et al.~(\cite{kdhscb97}) \\
5730 & 5250 & 4.2  & 4.6  & 0.20 & 0.22 &&& Th\'evenin~(\cite{t98}) \\
\\ \hline \\
5790 & 5260 & 4.32 & 4.51 & 0.20 & 0.23 &&&This paper \\ 
\\ \hline 
\end{tabular}
\end{table*}

\begin{table}
\caption[]{
Atmospheric abundances of $\alpha$\,Cen\,A \& B. The references are:
(1) Furenlid \& Meylan~(\cite{fm90}),         
(2) Neuforge-Verheecke \& Magain~(\cite{nm97}),
(3) Th\'evenin~(\cite{t98}).
}\label{tab:dataab}
\begin{tabular}{llllllllllllll} \hline \\
&\multicolumn{3}{c}{$\alpha$\,Cen\,A}&&\multicolumn{2}{c}{$\alpha$\,Cen\,B} \\
\\ \hline \\
C  &           &0.20\,(2) &0.14\,(1)&&         &0.28\,(2)\\
N  &           &0.30\,(2)\\
O  & 0.15\,(3) &0.21\,(2)&0.10\,(1)\\
Al & 0.30\,(3) &0.24\,(2)&0.23\,(1)&& 0.40\,(3)&0.24\,(2)\\
Si & 0.35\,(3) &0.27\,(2)&0.20\,(1)&& 0.45\,(3)&0.27\,(2)\\
Ca & 0.30\,(3) &0.22\,(2)&0.13\,(1)& &         &0.21\,(2)\\
Sc & 0.30\,(3) &0.25\,(2)&0.13\,(1)&&          &0.26\,(2)\\
Ti & 0.25\,(3) &0.25\,(2)&0.08\,(1)&&          &0.27\,(2)\\
V  & 0.30\,(3) &0.23\,(2)&0.08\,(1)&&          &0.32\,(2)\\
Cr & 0.25\,(3) &0.25\,(2)&0.13\,(1)&&          &0.27\,(2)\\
Mn &           &0.23\,(2)&0.25\,(1)&&          &0.26\,(2)\\
Fe & 0.20\,(3) &0.25\,(2)&0.12\,(1)&& 0.22\,(3)&0.24\,(2)\\
Co & 0.35\,(3) &0.28\,(2)&0.16\,(1)&&          &0.26\,(2)\\
Ni & 0.28\,(3) &0.30\,(2)&0.16\,(1)&&          &0.30\,(2)\\
Y  & 0.20\,(3) &0.20\,(2)&0.05\,(1)&&          &0.14\,(2)\\
Zr & 0.20\,(3) &0.17\,(2)&0.04\,(1)&&          &         \\
La & 0.40\,(3) & \\
Ce & 0.40\,(3) &\\
Nd & 0.35\,(3) &0.20\,(2)\\
Sm & 0.40\,(3) &\\
Eu  &          &0.15\,(2)\\
\\ \hline
\end{tabular}
\end{table}

\subsection{Spectroscopic data and atmospheric parameters}\label{sec:cp}
For both stars many spectroscopic data exist in the literature
with good signal to noise ratios because of their high magnitudes.
As Table~\ref{tab:datas} exhibits
rather scattered results we decided to reestimate
the effective temperatures, surface gravities and metallicities. For our
analysis we used the spectroscopic data published
by Chmielewski et al.~(\cite{cfcb92}) and 
Neuforge-Verheecke \& Magain~(\cite{nm97}), 
the atmosphere models of Bell et al.~(\cite{b76}) grid
and the basic technique of the curve of growth.

\subsubsection{Effective temperatures}
Table~\ref{tab:datas} presents effective temperatures compiled from 
the literature. Chmielewski et al.~(\cite{cfcb92})
derived the effective temperatures from the best fit of
the wings of the $\rm H_{\alpha}$ line.
 Recently Neuforge-Verheecke \& Magain~(\cite{nm97}) 
obtained the $T_{\rm eff}$ by forcing the abundances derived from 
FeI lines to be independent of their excitation potential.
They not only used the FeI lines to determine the effective
temperature of each star but also checked their results using the
$\rm H_{\alpha}$ line profiles.
In $\alpha$\,Cen\,A, the two methods lead to consistent results that are
also in good agreement with those of Chmielewski et al.~(\cite{cfcb92}),
although the two analyses made use of different model atmospheres.
In $\alpha$\,Cen\,B, the two methods lead to marginally discrepant
results
possibly due to Non-LTE effects and to the sensitivity of $\rm
H_{\alpha}$ to the treatment of convection.

We have reconsidered the fit of hydrogen lines
of Chmielewski et al.~(\cite{cfcb92}). We took into account
the overabundance of the $\alpha$\,Cen system which strengthens
the metallic lines
and displaces the pure wings of H lines; we also took into account
the relative strength
of metallic lines in $\alpha$\,Cen\,B, different from what is found in
solar-like profiles, because of the lower temperature. Consequently,
the fit proposed by
Chmielewski et al.~(\cite{cfcb92}) seems to overestimate the temperature
 and we displace it toward cooler values,
more importantly in the case of $\alpha$\,Cen\,B.
Our adopted effective temperatures are respectively $5790\pm30$\,K and 
$5260\pm50$\,K for $\alpha$\,Cen\,A \& B. 
These new proposed temperatures agree,
within the 1\,$\sigma$ error bars, with the FeI based temperature
determinations of Neuforge-Verheecke and Magain~(\cite{nm97}) and,
as quoted in Sect.~\ref{sec:p}, they are close to the
values based on Lejeune et al.~(\cite{lcb98}) calibrations.

\subsubsection{Surface gravities}\label{sec:sg}
Usually $\log g$, the logarithm of the surface gravity, is determined by
the detailed spectroscopic analysis
using the ionization equilibrium of iron. Table~\ref{tab:datas} shows
the results obtained by different authors.
We remind that the surface gravities from Neuforge-Verheecke \&
Magain~(\cite{nm97})
and from Chmielewski et al.~(\cite{cfcb92}) are very close.
Using the published equivalent
width (hereafter EW) and the error bars
of Neuforge-Verheecke \& Magain, we derive
by the basic curve of growth procedure $\log g_{\rm A}=4.32\pm0.05$ and
$\log g_{\rm B}=4.51\pm0.08$ for $\alpha$\,Cen\,A \& B, respectively.

\subsubsection{Chemical composition}
Furenlid \& Meylan~(\cite{fm90}) and Neuforge-Verheecke \&
Magain~(\cite{nm97})
 have found an average metal overabundance of around 0.2 dex.
As in Sect.~\ref{sec:sg}, using oscillator strengths,
microturbulence parameter, EW, published by
Neuforge-Verheecke \& Magain
and the iron curve of growth derived as
aforesaid, we get $\rm[\frac{Fe}H]_A=0.20\pm0.02$\,dex and
$\rm[\frac{Fe}H]_B=0.23\pm0.03$\,dex for $\alpha$\,Cen\,A \& B,
respectively;
the error bars are from Neuforge-Verheecke \& Magain.
We note that metallicities agree,
within the 1\,$\sigma$ error bars, with the values of
Neuforge-Verheecke \& Magain.

Table~\ref{tab:dataab} shows that C, N, O and Fe, 
the most abundant ``metals'' and electron donors, have uniform
overabundances of about $\sim0.2$\,dex.
In consequence, it is reasonable to construct models having a uniform 
abundance of all metals enhanced by a factor 1.7, corresponding to 0.2
dex compared to 
the Sun, and use opacities with solar mixture. On the main-sequence,
for stellar masses close to the
solar one, the differential segregation between the metals does not
significantly modify the models (Turcotte et al.
\cite{trmir98}) this permits, for the equation of state and opacity data,
to keep the ratios between ``metals'' to their initial value despite the
differential segregation by microscopic diffusion and gravitational
settling.

Table \ref{tab:datas} reveals a good agreement between the measurements
of the
lithium abundance. $\alpha$\,Cen\,A presents
a depletion close to the solar one ($\rm[Li]_{s\,\odot}=1.11$,
Grevesse \& Sauval~\cite{gs98}) while, in $\alpha$\,Cen\,B, the lithium
 is clearly more depleted.
The assumption of a common origin of the two stars eliminates
the possibility of a different initial lithium abundance
in the pre-stellar nebulae. Neither an error on temperatures nor on 
surface gravities in model atmospheres
can explain these differences with the solar depletion.

The situation is similar for \element[][9]{Be}.
Observations by Primas et al.~(\cite{pdp97})  indicate
that \element[][9]{Be} is depleted in  $\alpha$\,Cen\,B while 
$\alpha$\,Cen\,A exhibits a quasi-solar abundance.

\subsubsection{Rotation}
For $\alpha$\,Cen\,A, Boesgaard \& Hagen~(\cite{bh74})
estimated a rotational period 10\% larger
than the solar one. More recently Saar \& Osten~(\cite{so97}) measured
rotational velocities for $\alpha$\,Cen\,A \& B of respectively
$2.7\pm0.7$\,km\,s$^{-1}$ and $1.1\pm0.8$\,km\,s$^{-1}$, in good
agreement with other determinations. With the estimated radius of
$R_{\rm \alpha\,Cen\,A}\simeq 1.2R_\odot$ and
$R_{\rm \alpha\,Cen\,B}\simeq 0.91R_\odot$,
the periods of rotation are respectively
$P_{\rm \alpha\,Cen\,A}\simeq 22$\,d,
and $P_{\rm \alpha\,Cen\,B}\simeq 41$\,d.
They bracket the solar value. 

\subsection{Seismological observations}
Many attempts have been made to detect solar-like $p$-mode oscillations
in $\alpha$\,Cen\,A, from ground-based observations. 
They lead to controversial results.
Gelly et al.~(\cite{ggf86}) reported a
detection, but the claimed mean large separation between the oscillation
frequencies $\Delta\nu_0=165.5\,\mu$Hz was very large
and inconsistent with the theoretical value (Demarque et al.~\cite{dga86}).
Despite the use of more sensitive techniques,
Brown \& Gilliland~(\cite{bg90}) failed to detect any oscillation and gave an
upper limit for the amplitude of 0.7\,m\,s$^{-1}$, i.e. 2-3 times the solar one.
Pottasch et al.~(\cite{pbh92})
have  detected  oscillations in $\alpha$\,Cen\,A. They found a set of
regularly spaced peaks,  with a mean separation corresponding
to $\Delta\nu_0=110\,\mu$Hz, consistent with the known properties of this star,
but with an amplitude of oscillation larger than the upper
limit of Brown \&  Gilliland~(\cite{bg90}). Edmonds
\& Cram~(\cite{ec95}) did not detect unambiguously the $p$-mode
oscillations, but they found an evidence for almost the same periodicity
as that found by Pottasch et al.~(\cite{pbh92})
in the power spectrum. More recently,
Kjeldsen et al.~(\cite{kbfd99}) found tentative evidence for
$p$-mode oscillations in $\alpha$\,Cen\,A,  from equivalent width
measurements of the Balmer hydrogen lines. They proposed  four 
different possible
identifications of the eight observed frequency peaks in the power spectrum,
which correspond to sets of the large and small frequency
spacings $\Delta\nu_0$ and $\overline{\delta\nu}_{0,2}$ of
(106.94,\,12.30), (106.99,\,8.15), (100.77,\,11.70) and (100.77,\,6.42),
in $\mu$Hz.
The frequency resolution of the seismological observations will be 
much improved in a near future, up to 0.1 $\mu$Hz,  with several  
ground-based and   space projects.

\begin{table*}
\caption[]{ 
 All symbols have their ordinary meaning.
\{The brackets indicate values derived from basic formula.\}

Top, left panel: adopted observational data to be reached by the calibration.

Top, right panel: Observational constraints of the ``best'' model of
Guenther \& Demarque~(\cite{gd00}).

Bottom panel:
Characteristics
of $\alpha$\,Cen\,A \& B and Sun models computed with the same input physics.
The $\alpha$\,Cen\,A ($respt.$ $\alpha$\,Cen\,B)
models are named ``A..'' ($respt.$ ``B..'').
The first four rows recall some items of Table~\ref{tab:glob}: the ages (Myr),
the initial helium mass fractions,
the initial heavy element to hydrogen 
mass ratios, and the mixing-length parameters.
The next three rows give the effective temperatures in K,
the surface gravities and the metallicities. The six next rows present respectively
 the luminosities,
the total radii in solar units, the surface mass fractions of hydrogen and
helium, the ratios of heavy element to hydrogen and the lithium abundances.
In the next rows
$R_{\rm cz}$ and $T_{\rm cz}$ are respectively the radius and the temperature
(in M\,K) at the base of the external convection zone,
$R_{\rm co}$ is the radius of the convective core (including overshoot for model $\rm A_{ov}$ \& $\rm B_{ov}$). At center,
$T_{\rm c}$, $\rho_{\rm c}$, $X_{\rm c}$, $Y_{\rm c}$ are respectively the
temperature (in M\,K), the density (in g\,cm$^{-3}$), the
hydrogen and the helium mass fractions.
}\label{tab:mod}
\vbox{
\centerline{
\begin{tabular}{llllllllllllllll}  \hline \\
&\multicolumn{3}{c}{Models {\scriptsize BV, ov \& CM}}&&\multicolumn{3}{c}{Models {\scriptsize GD}}& \\
                   &&$\alpha$\,Cen\,A&$\alpha$\,Cen\,B&&&$\alpha$\,Cen\,A&$\alpha$\,Cen\,B\\ \\ \hline \\
$M/M_\odot$        &&$1.16\pm0.031$   &$0.97\pm0.032$   &&&$1.1015\pm0.008$ &$0.9159\pm0.007$\\
$T_{\rm eff}$      &&$5790\pm30$\,K   &$5260\pm50$\,K   &&&$5770\pm50$\,K	 &$5300\pm50$\,K\\
$\log g$           &&$4.32\pm0.05$    &$4.51\pm0.08$    &&&$\{4.28\pm0.02$	 &$4.54\pm0.03\}$  \\
$\rm [\frac{Fe}H]$ &&$0.20\pm0.02$    &$0.23\pm0.03$    &&&$0.22\pm0.02$	 &$0.26\pm0.04$  \\
$L/L_\odot$       &&$\{1.534\pm0.103$&$0.564\pm0.064\}$&&&$1.572\pm0.135$  &$0.509\pm0.06$\\
\end{tabular}
}
\vspace{0.5truecm}
\centerline{
\begin{tabular}{llllllllllllllllllll}  \hline \\
&\multicolumn{11}{c}{Models with $\rm MLT_{BV}$ theory}&\multicolumn{5}{c}{Models with $\rm MLT_{CM}$ theory} \\
			  &$\rm A_{BV}$ 	 &$\rm B_{BV}$	  &&$\rm A_{ov}$	  &$\rm B_{ov}$   &&$\rm A_{GD}$	 &$\rm B_{GD}$   &&$\odot_{\rm BV}$&&& $\rm A_{CM}$&$\rm B_{CM}$&&$\odot_{\rm CM}$\\
\\ \hline \\
$t_{\rm\alpha\,Cen}$ &\multicolumn{2}{c}{2710}  &&\multicolumn{2}{c}{3530}    &&\multicolumn{2}{c}{5640}  &&4685  &&&\multicolumn{2}{c}{4086}  &&4685  \\
$Y_{\rm i}$    &\multicolumn{2}{c}{0.284} &&\multicolumn{2}{c}{0.279}   &&\multicolumn{2}{c}{0.300} &&0.274 &&&\multicolumn{2}{c}{0.271} &&0.274 \\
$(\frac ZX)_{\rm i}$&\multicolumn{2}{c}{0.0443}&&\multicolumn{2}{c}{0.0450}&&\multicolumn{2}{c}{0.0480}&&0.0279&&&\multicolumn{2}{c}{0.0450}&&0.0279\\
$\alpha$ 		  &1.53  &1.57	  &&1.64  &1.66	 &&1.86	 &1.97     && 1.93&&&0.96  &0.99   && 1.03\\
 \\
$T_{\rm eff}$		  &5795  &5269    &&5795  &5269  &&5759  &5332  &&5778    &&&5794  &5268   &&5778\\
$\log g$		  &4.324 &4.508   &&4.316 &4.509 &&4.286 &4.523 && 4.438  &&&4.324 &4.522  && 4.438\\
$\rm [\frac{Fe}H]$        &0.20 &0.23    &&0.20  &0.23  &&0.22  &0.25  && 0.000  &&&0.20  &0.23   && 0.000\\
\\
$L/L_\odot$		  &1.527 &0.571   &&1.555 &0.569 &&1.565 &0.547 && 1.000  &&&1.526 &0.552  && 1.000\\
$R/R_\odot$		  &1.228 &0.909   &&1.239 &0.907 &&1.259 &0.867 && 1.000  &&&1.228 &0.893  && 1.000\\
$X_{\rm s}$		  &0.722 &0.701   &&0.729 &0.708 &&0.711 &0.695 && 0.737  &&&0.735 &0.717  && 0.737\\
$Y_{\rm s}$		  &0.250 &0.270   &&0.243 &0.262 &&0.260 &0.275 && 0.245  &&&0.236 &0.253  && 0.245\\
$\rm [Li]_{\rm s}$	  &2.72  &1.10    &&2.54  &0.496 &&1.94  &-3.32 && 2.46   &&&2.33  &-0.54  && 2.46 \\
$(\frac ZX)_{\rm s}$      &0.0384&0.0418  &&0.0385&0.0419&&0.0410&0.0433&& 0.0245 &&&0.0386&0.0418 && 0.0245\\
\\
$R_{\rm cz}/R_\star$	  &0.776 &0.698   &&0.758 &0.692 &&0.707 &0.675 && 0.715  &&&0.741 &0.685  && 0.715\\
$T_{\rm cz}$		  &1.503 &2.585   &&1.644 &2.653 &&2.020 &2.846 && 2.178  &&&1.801 &2.761  && 2.178\\
\\
$R_{\rm co}/R_\star$	  &0.026 &        &&0.054 &      &&0.061 &      &&        &&&0.038 &       && \\
$T_{\rm c}$		  &17.00 &13.66   &&17.46 &13.76 &&19.43 &14.26 && 15.77  &&&17.74 &13.71  && 15.76\\
$\rho_{\rm c}$       	  &126.0 &96.28   &&125.5 &100.8 &&172.0 &123.7 && 155.9  &&&146.8 &102.4  && 155.9\\
$X_{\rm c}$		  &0.374 &0.540   &&0.387 &0.510 &&0.178 &0.400 && 0.330  &&&0.283 &0.497  && 0.330\\
$Y_{\rm c}$		  &0.594 &0.428   &&0.580 &0.458 &&0.788 &0.566 && 0.649  &&&0.684 &0.470  && 0.649\\
\\ \hline
\end{tabular}
}
}
\end{table*}

\subsection{Adopted observables constraining the models}
To constrain the models at present day,
contrarily to previous works, we prefer to use
the spectroscopic gravities we derived consistently from
effective temperature and metallicities
instead of the luminosity derived from the photometry,
bolometric correction and parallax.
Table~\ref{tab:mod} gives
the effective temperatures, gravities and metallicities
we selected to constrain the models.
The masses are from Pourbaix et al.~(\cite{pnn99}).
We emphasize that the parallax
is required to derive the masses but neither
the effective temperatures nor the gravities.

While this work was under investigation, Guenther \& Demarque~(\cite{gd00})
published new calibrations of the $\alpha$\,Cen binary system.
As a matter of
comparison we have calibrated the $\alpha$\,Cen binary system with
their constraints. 
Table~\ref{tab:mod} presents the observable constraints of the
 ``best'' model of Guenther \& Demarque~(\cite{gd00}).
They significantly differ from ours, though correspondingly
to almost the same mass ratios.
The trigonometrical parallax of S\"oderhjelm~(\cite{s00}) is larger
than the orbital parallax of Pourbaix et al.~(\cite{pnn99}) and 
 leads to smaller masses and larger ($respt.$ smaller) luminosity
 for $\alpha$\,Cen\,A  ($respt.$ $\alpha$\,Cen\,B).

\section{The method of calibration}\label{sec:chi2}
The calibration of a binary system is based on the adjustment
of stellar modeling parameters
 to observational data at the age of the system. For a given mass,
fixing the physics,
the effective temperature, surface gravity and metallicity
of a model have the formal dependences with respect to modeling parameters:
\begin{eqnarray*}
T_{\rm eff}(\star)_{\rm mod}&=&T_{\rm eff}\left(t_\star;Y_{\rm i},{\rm[\frac{Fe}H]_i,\alpha_\star}\right),\\
\log g(\star)_{\rm mod}&=&\log g\left(t_\star;Y_{\rm i},{\rm[\frac{Fe}H]_i},\alpha_\star\right),\\
{\rm[\frac{Fe}H]_{\rm s}}(\star)_{\rm mod}&=&{\rm[\frac{Fe}H]_{\rm s}}\left(t_\star;Y_{\rm i},{\rm[\frac{Fe}H]_i},\alpha_\star\right).
\end{eqnarray*}
The basic idea of the $\chi^2$ fitting has been developed by
Lastennet et al.~(\cite{lvlo99}). To find a set of modeling parameters: 
\[\wp_{\rm \alpha\,Cen} \equiv\left\{t_{\rm \alpha\,Cen},
Y_{\rm i}, \rm[\frac{Fe}H]_i,\alpha_{\rm A},\alpha_{\rm B}
\right\},\]
leading to observables as close as possible to the observations
$T_{\rm eff}(\star)$, $\log g(\star)$ and $\rm[\frac{Fe}H]_{\rm s}(\star)$,
we minimize the
$\chi^2(t_\star,Y_{\rm i},{\rm[\frac{Fe}H]_i},\alpha_{\rm A},\alpha_{\rm B})$
functional defined as:
\begin{eqnarray}
\chi^2&=&
\sum_{\star=A,B}\left(\frac{T_{\rm eff}(\star)_{\rm mod}-T_{\rm eff}(\star)}
{\sigma\left(T_{\rm eff}(\star)\right)}\right)^2 + \nonumber\\
&+&\left(\frac{\log g(\star)_{\rm mod}-\log g(\star)}
{\sigma\left(\log g(\star)\right)}\right)^2 + \label{eq:chi2}\\
&+&\left(\frac{{\rm [\frac{Fe}H]_s}(\star)_{\rm mod}-{\rm [\frac{Fe}H]_{\rm s}}(\star)}
{\sigma\left({\rm [\frac{Fe}H]_{\rm s}}(\star)\right)}\right)^2, \nonumber
\end{eqnarray}
where $\sigma\left(T_{\rm eff}(\star)\right)$,
$\sigma\left(\log g(\star)\right)$ and
$\sigma\left({\rm [\frac{Fe}H]_{\rm s}}(\star)\right)$ are the
uncertainties associated to each star.
For a grid of modeling parameters $\wp$, 
we have computed main-sequence evolution of
models with $\alpha$\,Cen\,A \& B masses. Then
the $\chi^2$ was computed using Eq.~(\ref{eq:chi2}) in a refined grid 
obtained by interpolations. We kept for the solution
the ``best'' $\wp=\wp_{\rm \alpha\,Cen}$ which corresponds
to the $\chi^2_{\rm min}$.
These best modeling parameters are used to compute
 models of $\alpha$\,Cen\,A \& B
including pre main-sequence and their frequencies. We do not further
attempt neither to improve nor to investigate the stability of the solution
via the steepest descent method
(Noels et al.~\cite{ngmnbl91}, Morel et al.~\cite{mmpb00}).
Table~\ref{tab:glob}
shows the confidence limits of modeling parameters
of models computed in this paper.
The confidence limits of each modeling parameter, the other being fixed, 
correspond to the maximum/minimum values it can reach, in order that the
generated models fit the observable targets within their error bars. 

\section{Evolutionary models}\label{sec:comp}
Basically the physics employed for the calculation of models is the same as in
Morel et al.~(\cite{mpb97}). The ordinary assumptions of stellar modeling are
made, i.e. spherical symmetry, no rotation, no magnetic field, no mass loss.
It has been already claimed, see Morel et al.~(\cite{mpb00}),
that stellar models of about one solar mass,
computed with the same physics but initialized either
on pre main-sequence or at zero-age homogeneous main-sequence
are almost identical after a few $10^7$\,years,
a small quantity with respect to the expected age of the $\alpha$\,Cen system.  
To save computations, along the search of a solution with the
$\chi^2$ minimization, we have initialized each evolution 
with a homogeneous zero-age main-sequence model. 
The models presented in Table~\ref{tab:mod}
 include the pre main-sequence evolution. They are
initialized with homogeneous zero-age stellar model in quasi-static contraction
with a central temperature
close to the onset of the deuteron burning, i.e. $T_{\rm c}\sim0.5$\,MK.

\subsection{Nuclear and diffusion network.}
We have taken into account the 
most important nuclear reactions of PP+CNO cycles (Clayton,~\cite {c68}).
The relevant nuclear reaction rates
are taken from the NACRE compilation (Angulo et al.~\cite{aar99}) with the
reaction
$\element[][7]{Be}(e^-,\nu\gamma)\element[][7]{Li}$ taken from the
compilation of
Adelberger et al.~(\cite{aab98}). Weak screening (Salpeter~\cite{s54}) is assumed.
We have used the meteoritic value (Grevesse \& Sauval~\cite{gs98}) for the
initial lithium abundance,
$\left[\element[][]{Li}\right]_{\rm i}=3.31\pm0.04$.
For the calculations of the depletion, lithium
is assumed to be in its most abundant isotope \element[][7]{Li} form.
The initial abundance of each isotope is derived from
isotopic fractions and initial values of
$Y\equiv \element[][3]{He}+\element[][4]{He}$ and $Z$ in order to fulfill 
the basic relationship $X+Y+Z\equiv1$ with
$X\equiv \element[][1]{H}+\element[][2]{H}$.
For the models computed for the $\chi^2$ fitting, the isotopes
 \element[][2]{H},
\element[][7]{Li}, \element[][7]{Be} are set at equilibrium.

Microscopic diffusion is described by the simplified formalism of 
Michaud \& Proffitt~(\cite{mp93}) with heavy elements as
trace elements. We have neglected the radiative accelerations as they
amount only to a tiny fraction of gravity 
in the radiative part for stars with masses close to the solar one
(Turcotte et al.~\cite{trmir98}).
We assume that changes of $Z$, as a whole, describe the
changes of metals and we use the approximation:
\begin{eqnarray}\label{eq:fesh}
\mathrm{[\frac{Fe}H]} &\equiv&\log(\frac{Z_{\rm Fe}}Z)+\log(\frac ZX)-
\log(\frac{Z_{\rm Fe}}X)_\odot\simeq \nonumber \\
&\simeq&\log(\frac ZX) - \log(\frac ZX)_\odot \nonumber
\end{eqnarray}
with $\frac{Z_{\rm Fe}}Z$ is the iron mass fraction within $Z$.

\subsection{Equation of state, opacities, convection and
atmosphere.}\label{sec:mix}
We have used the CEFF equation of state (Christensen-Dalsgaard \& D\"appen
\cite{cd92})
and the opacities of Iglesias \& Rogers~(\cite{ir96})
complemented at low temperatures by
Alexander \& Ferguson~(\cite{af94}) opacities
for the solar mixture of Grevesse \& Noels~(\cite{gn93}). 
We have not taken into account
the changes of abundance ratios between the metals within $Z$ due to diffusion;
for stellar masses close to the solar one they do not really affect
the structure of models (Turcotte et al.~\cite{trmir98}).

In the convection zones the temperature gradient is
computed according to either $\rm MLT_{BV}$ or $\rm MLT_{CM}$
convection theories.
The mixing-length is defined as $l\equiv \alpha H_{\rm p}$,
where $H_{\rm p}$ is the pressure scale height.
The convection zones are mixed via
a strong full mixing turbulent diffusion 
coefficient $d_{\rm m}=\rm 10^{13}\,cm^2\,s^{-1}$ which produces a homogeneous
composition (Morel~\cite{m97}).

At the end of the pre main-sequence both components,
have for a few million years,
 a temporary convective core.
For $\alpha$\,Cen\,A, slightly more massive than the Sun, a second
convective core is formed during the main-sequence due to
the onset of the CNO burning
(e.g. Guenther \& Demarque~\cite{gd00}). Following
the prescriptions of Schaller et al.~(\cite{ssmm92}) we have calibrated models
with overshooting of
convective cores over the distance
$O_{\rm v}=0.2\min(H_{\rm p},R_{\rm co})$ where
$R_{\rm co}$ is the core radius.

The atmosphere is restored using a grid of $T(\tau,T_{\rm eff},g)$ laws,
provided by Cayrel~(\cite{c00}),
($\tau$ is the Rosseland optical depth) derived
from atmosphere models with the solar mixture of
Grevesse \& Noels~(\cite{gn93}) and
metallicity $\rm [\frac{Fe}H]=0.2$. The atmosphere models were computed 
with the Kurucz~(\cite{k91}) ATLAS12 package.
The connection with the envelope is made at the optical
depth $\tau_{\rm b} = 20$ where the diffusion approximation for radiative
transfer becomes valid (Morel et al.~\cite{mvpbccl94}).
A smooth connection of the
gradients is insured between the uppermost layers of the envelope
and the optically thick convective bottom of the atmosphere.
It is an important requirement for the
calculation of eigenmode frequencies.
The radius $R_\star$ of any model
is taken at the optical depth $\tau_\star$ where
$T(\tau_\star)=T_{\rm eff}$. Typically, $\tau_\star$ increases from
$\tau_\star\sim0.43$ in the initial pre main-sequence model,
until $\tau_\star\sim0.53$ at the present time.
The mass of the star $M_\star$ is defined as the mass enclosed in
the sphere of radius $R_\star$.
The external boundary is located at the optical
depth $\tau_{\rm ext}=10^{-4}$, where the density is fixed to its value in
the atmosphere model $\rho(\tau_{\rm ext})=2.22\,10^{-9}$\,g\,cm$^{-3}$.
To simplify, the chemical composition within the atmosphere models is assumed
to be unaffected by the diffusion.

\subsection{Numerics.}
Models have been computed using the CESAM code (Morel~\cite{m97}).
The numerical schemes are fully implicit and their accuracy
is of the first order for the time and third order for the space.
Each model is described by about 600 mass shells, this number
 increases up to 2100 for the models used in seismological analysis.
Evolutions are described by about 80 models. About half of them
concerns the pre main-sequence evolution.

\begin{figure*}
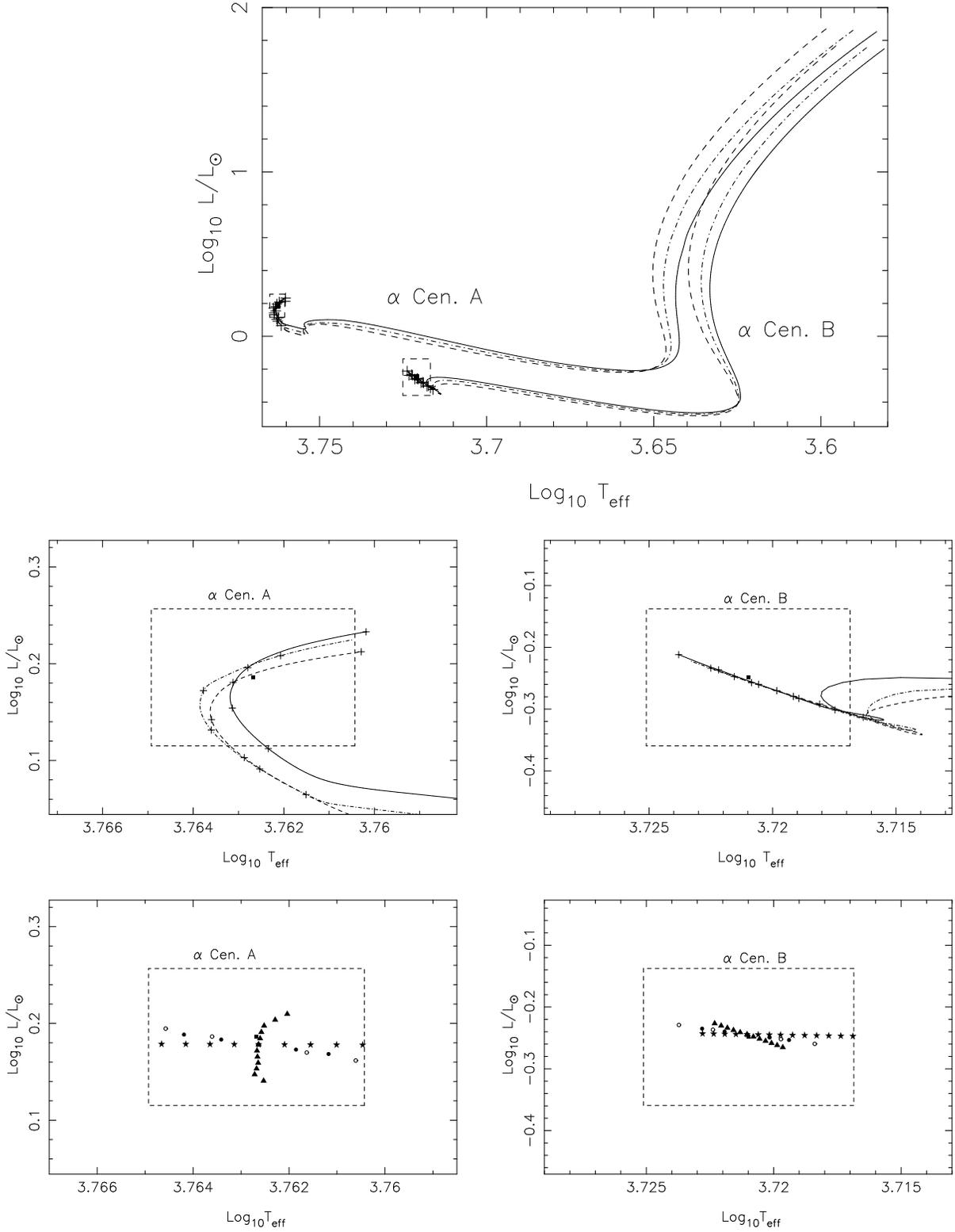

\vbox{
\centerline{
\psfig{figure=10088.f6,height=8.5cm,angle=270}
}
\vspace{0.5truecm}
\hbox{\psfig{figure=10088.f7,height=5.5cm,angle=270}
\hspace{0.5truecm}
\psfig{figure=10088.f8,height=5.5cm,angle=270}}
\vspace{0.5truecm}
\hbox{\psfig{figure=10088.f9,height=5.5cm,angle=270}
\hspace{0.5truecm}
\psfig{figure=10088.f10,height=5.5cm,angle=270}}
}
\caption{
Evolutionary tracks in the H-R diagram of models $\rm A_{BV}$,
$\rm B_{BV}$ (full),
$\rm A_{CM}$, $\rm B_{CM}$ (dashed) and $\rm A_{ov}$, $\rm B_{ov}$ (dot-dash-dot).
Dashed rectangles delimit the uncertainty domains.
Top panel: full tracks from PMS. The stellar evolution sequences are
initialized on the pre main-sequence soon after the deuteron ignition.
The ``+'' denote 1\,Gyr time intervals along the evolutionary tracks.
Middle left and right panels: enlargements around the observed
$\alpha$\,Cen\,A, \& B loci.
Bottom left and right panels: loci of $\alpha$\,Cen\,A \& B models
computed with the $\rm MLT_{BV}$ convection theory and
modeling parameters within the confidence domains presented in
Table~\ref{tab:mod}; full triangle: $t_{\rm\alpha\,Cen}$
changes only, full star: $\alpha$ changes only, empty circle:
$Y_{\rm i}$ changes only  and full dot: $\rm[\frac{Fe}H]_i$ changes only.
}\label{fig:AB}
\end{figure*}

\begin{table*}
\caption[]{
Partial derivatives of observables
$\log T_{\rm eff\,A}$,
$\log\frac {L_{\rm A}}{L_\odot}$,
$\rm\left[\frac {Fe}H\right]_A$,
$\log T_{\rm eff\,B}$,
$\rm\alpha_B$, 
$\log\frac {L_{\rm B}}{L_\odot}$ and
$\rm\left[\frac {Fe}H\right]_B$
with respect to modeling parameters
$t_{\rm \alpha\,Cen}$(Myr),
$\rm\alpha_A$, $\rm\alpha_B$,
$Y_{\rm i}$ and
$\rm\left[\frac {Fe}H\right]_i$ of models $\rm A_{BV}$ \& $\rm B_{BV}$.
}\label{tab:jacob}
\begin{tabular}{lllllll} \\  \hline \\
&$\partial\log T_{\rm eff\,A}$
&$\partial\log\frac {L_{\rm A}}{L_\odot}$
&$\rm \partial\left[\frac {Fe}H\right]_A$
&$\partial\log T_{\rm eff\,B}$
&$\partial\log\frac {L_{\rm B}}{L_\odot}$
&$\rm \partial\left[\frac {Fe}H\right]_i$ \\
 \\ \hline \\
$\partial t_{\rm \alpha\,Cen}$         &$-2.15\,10^{-7}$&$4.16\,10^{-5}$&$-2.30\,10^{-5}$&$1.63\,10^{-6}$ &$2.32\,10^{-5}$ &$-9.24\,10^{-6}$\\ \\
$\partial\rm\alpha_A$                   &$0.0380$        &$8.23\,10^{-3}$&$0.0541$        &$0.$            &$0.$            &$0.$            \\ \\
$\partial\rm\alpha_B$                   &$0.$            &$0.$           &$0.$            &$0.0306$        &$0.0213$        &$5.67\,10^{-3}$ \\ \\
$\partial Y_{\rm i}$                    &$0.469$         &$3.85$         &$-0.911$        &$0.621$         &$3.55$          &$-0.185 $\\ \\
$\rm \partial\left[\frac {Fe}H\right]_i$&$-2.12\,10^{-3}$ &$-0.014$       &$0.024$         &$-2.40\,10^{-3}$&$-0.0129$       &$0.0205$\\
\\ \hline
\end{tabular}
\end{table*}

\section{Results}\label{sec:res}
For different sets of modeling parameters
we have computed evolutionary tracks
using the convection theories of B\"ohm-Vitense~(\cite{b58},
$\rm MLT_{BV}$) and Canuto \& Mazitelli (\cite{cm91},~\cite{cm92},
$\rm MLT_{CM}$). Using the
$\chi^2$ minimization described in Sect.~\ref{sec:chi2},
we have selected the best
fits taking into account the accuracy of the observational constraints.
Tables~\ref{tab:glob} and \ref{tab:mod} (bottom panel)
present the characteristics
of $\alpha$\,Cen\,A \& B and Sun models computed with the same input physics.
Figure~\ref{fig:AB} presents the evolutionary tracks in the 
$\log L/L_\odot$ versus $\log T_{\rm eff}$ H-R diagram.
We also performed $\chi^2$ minimization with the constraint
$\rm \alpha_A\equiv\alpha_B$, but, for brevity sake,
we do not report the results as they do not significantly differ
from those obtained without this constraint.
For all $\alpha$\,Cen\,A models a convective core, caused by an enhancement of
the CNO nuclear energy generation, appears as soon as
the central temperature becomes slightly larger than 17\,M\,K and the mass
fraction of hydrogen $X_{\rm c}\sol 0.37$.
Figure~\ref{fig:AB}, Tables~\ref{tab:glob} and~\ref{tab:mod} clearly show
that various sets of modeling parameters
allow to verify the observational constraints within the error bars.
In absence of seismological observations
there is no criterion available to discriminate between models.
Tables~\ref{tab:glob} and~\ref{tab:mod} show that, within the confidence domains,
the values of the mixing length parameter of the two
components are very close.

\subsection{$\rm MLT_{BV}$ models.}
Models $\rm A_{BV}$ \& $\rm B_{BV}$ have an age
$t_{\rm \alpha\,Cen}=2710$\,Myr compatible with 
the estimate of Pourbaix et al.~(\cite{pnn99}). The values
 of the convection parameters are almost equal
($\alpha_{\rm A}\simeq\alpha_{\rm B}$) and close to the value
derived if they are
forced to be identical in the $\chi^2$ minimization. They are quite close to
the value obtained by Noels et al.~(\cite{ngmnbl91}) but differ notably
from the estimate of  Pourbaix et al.~(\cite{pnn99}) though remaining
within their large
confidence domains. They are smaller than the solar value $\alpha_\odot$. In 
$\rm \alpha\,Cen\,A$ a convective core 
is  formed at time $t\sim2.6$\,Gyr a few Myr just before
$t_{\rm \alpha\,Cen}$.
Figure~\ref{fig:AB}, bottom panels, shows the loci in the H-R diagram
reached by the $\rm A_{BV}$ \& $\rm B_{BV}$ models when one
modeling parameter changes within its uncertainty domain,
the other modeling parameters being fixed.
The limited extents of the permitted solutions
 are consequences of limits in metallicity.
Table~\ref{tab:jacob} shows the partial derivatives of the observed parameters
with respect to the modeling ones.
Our results are similar to those of Brown et al.~(\cite{bcwg94}), except
for the derivatives with respect to the age which are very
dependent on the exact
evolutionary stage on the main sequence.
As already noticed by Guenther \& Demarque~(\cite{gd00}),
the luminosities are very
sensitive to the helium mass fraction (see the large values of their
derivatives with respect to $Y_{\rm i}$ in Table~\ref{tab:jacob}),
which in turn gives a narrow
confidence level for the initial helium abundance (see Table~\ref{tab:glob}).

For both components
the surface lithium depletion is not sufficient enough to fit
the observed values.

\medskip

Models $\rm A_{ov}$ \& $\rm B_{ov}$ with overshooting of convective cores
have an age
significantly larger than models $\rm A_{BV}$ \& $\rm B_{BV}$. Indeed their outer
convective zones penetrate deeper.
For model $\rm B_{ov}$ this depletion is marginally compatible with the
observation, while it is not large enough for model $\rm A_{ov}$. All
these differences with models $\rm A_{BV}$ \& $\rm B_{BV}$, of course, result
 from the overshooting of convective cores 
but in a roundabout way via the $\chi^2$ minimization which adjusts
the modeling parameters as a whole.
The convective core of model $\rm A_{ov}$ is formed at time $t\sim3.16$\,Gyr.

\medskip

Models $\rm A_{GD}$ \& $\rm B_{GD}$ are calibrated according to the
Guenther \& Demarque~(\cite{gd00}) observational constraints and masses.
The age, initial helium mass fraction
and metallicity  and the convection parameters
are significantly larger than those derived using
the mass values of Pourbaix et al.~(\cite{pnn99}). We have obtained
similar modeling parameters from
the $\chi^2$ minimization using models with  
Guenther \& Demarque~(\cite{gd00}) masses and
our observational constraints. Therefore,
the differences between the modeling parameters of
models $\rm A_{BV}$ \& $\rm B_{BV}$ and
$\rm A_{GD}$ \& $\rm B_{GD}$ mainly result from the mass differences,
but also from the disparity of
observing targets, although in a less extent.
For models $\rm A_{GD}$ \& $\rm B_{GD}$
the convection parameters are close to the solar value. 
In the model $\rm A_{GD}$ the convective core is formed at the time $t\sim3.7$\,Gyr,
larger than in other $\rm MLT_{BV}$ models and
the lithium depletion at the surface is marginally compatible
with the observed value.
The value of the central hydrogen mass fraction is
half that obtained in the other $\rm MLT_{BV}$ models.
In model $\rm B_{GD}$ the lithium depletion is compatible with the observation.

\subsection{$\rm MLT_{CM}$ models.}
In agreement with Canuto \& Mazitelli~(\cite{cm91}, \cite{cm92})
convection theory,
the convection parameters of models $\rm A_{CM}$ \& $\rm B_{CM}$ are
close to unity and close to the
solar value. The age is slightly smaller than the solar one
but larger than that of models $\rm A_{BV}$ \& $\rm B_{BV}$;
the outer convection zone is deeper.
According to the observations, the lithium of
model $\rm B_{CM}$ is almost totally depleted but practically
no depletion occurred in models $\rm A_{CM}$ and the solar one.
The convective core of model $\rm A_{CM}$ is formed at time $t\sim3.9$\,Gyr.

\begin{table}
\caption[]{
 Theoretical global asymptotic characteristics of the low degree
$p$-mode  and $g$-mode spectrum
of the star $\alpha$\,Cen\,A \& B. The quantities $\nu_{00}$,
$\Delta\nu_0$
$\overline{\delta\nu}_{0,2}$ and  $\overline{\delta\nu}_{1,3}$ ($n_0=21$),
describing the $p$-mode oscillations,
are given in $\mu$Hz - see definitions Sect.~\ref{sec:pmode}.
The characteristic $g$-mode period
$P_0$ is given in mn. The lower panel presents the variations
$\Delta[\Delta\nu_0]$,$\Delta[\overline{\delta\nu}_{0,2}]$,
$\Delta[\overline{\delta\nu}_{1,3}]$ and $\Delta[P_0]$  for the models
computed with
the $\rm MLT_{BV}$ theory and extreme modeling parameters
$\alpha$, $t_{\rm\alpha\,Cen}$, $Y_{\rm i}$ and $\rm [\frac{Fe}H]_i$ within the
confidence domains presented in Table~\ref{tab:glob}.
}\label{tab:seis}
\begin{tabular}{llllll}  \\ \hline \\
   &$\nu_{0 0}$ &$\Delta\nu_0$ &  $\overline{\delta\nu}_{0,2}$  &
$\overline{\delta\nu}_{1,3}$&$P_0$\\
   \\ \hline \\
$\rm A_{BV}$ &  2391.0  & 108.1 &8.9    &  14.7   &   44.8\\
$\rm A_{ov}$ &  2365.9  & 106.8 &9.1   &  13.6    &   56.3\\
$\rm A_{CM}$ &  2397.3  & 108.1 &7.5   &  12.8    &   44.3\\
$\rm A_{GD}$ &  2256.9  & 101.7 &4.4   &  9.7     &   49.2\\
\\
$\rm B_{BV}$ &  3456.8  & 154.0 &12.5    &  20.6   &   55.5\\
$\rm B_{ov}$ &  3467.2  & 154.3 &12.0   &  19.8    &   52.1\\
$\rm B_{CM}$ &  3546.4  & 157.4 &11.7   &  19.5    &   50.9\\
$\rm B_{GD}$ &  3603.7  & 160.2 &10.7   &  18.4    &   43.6\\
\\ 
\end{tabular}
\begin{tabular}{llllll}   \hline \\
       &  $\Delta[\Delta\nu_0]$
&$\Delta[\overline{\delta\nu}_{0,2}]$&$\Delta[\overline{\delta\nu}_{1,3}]$
& $\Delta[P_0]$  \\ \\ \hline \\
$\rm A_{BV}$\\
$\Delta\alpha_{\rm A}=0.12$& 2.9   &0.1     &   0.3      &    0.4\\
$\Delta t_{\rm\alpha\,Cen}$=1300    &   -9.7     &-1.7     & -3.1       &   3.1\\
$\Delta Y_{\rm i}$=0.004 &  -0.9   &-0.1     & -0.3       &   3.6\\
$\Delta\rm [\frac{Fe}H]_i=0.004$&0.2 &0.01     & 0.04       &0.2\\
\\
$\rm B_{BV}$  \\
$\Delta\alpha_{\rm B}=0.14$& 3.6   & 0.1      &   0.2     &    0.2\\
$\Delta t_{\rm\alpha\,Cen}$=1300 &   -5.4   &-1.2     & -1.6       &   -8.2\\
$\Delta Y_{\rm i}$=0.004&  -1.1   &-0.1     & -0.1       &   0.9\\
$\Delta\rm [\frac{Fe}H]_i=0.004$&0.2 &0.01     & 0.01       &0.1\\
\\ \hline \\
\end{tabular}
\end{table}

\begin{table*}
\caption[]{
 Low degree frequencies for $\alpha$\,Cen\,A and B.
}\label{tab:seisab}
\begin{tabular}{llllllllllllll}  \hline \\
&\multicolumn{6}{c}{$\alpha$\,Cen\,A, model $\rm A_{BV}$} 
&\multicolumn{4}{c}{$\alpha$\,Cen\,B, model $\rm B_{BV}$}\\
\\ \hline \\
   $n$&&       $\ell= 0$      &$\ell= 1$     &$\ell= 2$  & $\ell= 3$&&
	       $\ell= 0$      &$\ell= 1$     &$\ell= 2$  & $\ell= 3$\\
\\ \hline \\
    1&&         216.50&         239.04&         332.72&         354.66&&
                280.33&         313.25&         404.94&         460.11\\
    2&&         330.58&         357.01&         414.45&         450.61&&
                439.72&         504.26&         575.00&         635.19\\
    3&&         433.74&         473.41&         525.29&         567.19&&
                594.94&         674.45&         748.46&         810.92\\
    4&&         539.84&         589.81&         640.82&         683.00&&
                765.25&         845.17&         916.83&         980.26\\
    5&&         651.62&         703.50&         754.02&         797.53&&
                932.81&        1013.63&        1085.89&        1148.42\\
    6&&         763.65&         817.42&         867.89&         911.05&&
               1101.83&        1179.48&        1251.72&        1315.75\\
    7&&         876.96&         929.96&         980.65&        1024.62&&
               1268.49&        1346.02&        1416.99&        1480.67\\
    8&&         989.90&        1043.11&        1093.39&        1136.78&&
               1433.70&        1510.26&        1581.42&        1645.03\\
    9&&        1102.65&        1154.54&        1204.33&        1247.43&&
               1598.27&        1673.35&        1742.83&        1806.05\\
   10&&        1213.79&        1264.47&        1312.49&        1354.49&&
               1759.39&        1833.61&        1901.74&        1963.56\\
   11&&        1321.79&        1370.95&        1418.40&        1460.41&&
               1917.69&        1989.93&        2057.43&        2119.37\\
   12&&        1427.80&        1476.58&        1523.87&        1566.71&&
               2073.10&        2145.00&        2212.06&        2274.53\\
   13&&        1533.43&        1583.08&        1631.25&        1674.60&&
               2227.40&        2299.89&        2367.70&        2430.43\\
   14&&        1640.95&        1690.78&        1738.77&        1782.29&&
               2382.75&        2455.04&        2522.86&        2585.94\\
   15&&        1748.45&        1798.25&        1845.79&        1888.78&&
               2537.58&        2609.85&        2677.04&        2739.71\\
   16&&        1855.24&        1904.34&        1951.62&        1994.97&&
               2691.18&        2762.91&        2830.18&        2892.90\\
   17&&        1961.01&        2010.40&        2057.91&        2101.70&&
               2843.92&        2915.35&        2982.68&        3046.16\\
   18&&        2067.20&        2117.07&        2165.32&        2209.86&&
               2996.13&        3068.24&        3136.07&        3200.00\\
   19&&        2174.62&        2225.10&        2273.55&        2318.42&&
               3149.16&        3221.60&        3290.20&        3354.77\\
   20&&        2282.71&        2333.49&        2382.26&        2427.33&&
               3303.04&        3375.80&        3444.50&        3509.53\\
   21&&        2391.27&        2442.08&        2490.76&        2536.02&&
               3456.99&        3530.11&        3599.13&        3664.29\\
   22&&        2499.56&        2550.55&        2599.41&        2644.81&&
               3611.26&        3684.29&        3753.67&        3819.43\\
   23&&        2608.02&        2659.04&        2708.02&        2753.86&&
               3765.55&        3838.97&        3908.55&        3974.76\\
   24&&        2716.48&        2767.90&        2817.19&        2863.25&&
               3920.10&        3993.88&        4064.07&        4130.66\\
   25&&        2825.46&        2877.03&        2926.58&        2973.04&&
               4075.36&        4149.28&        4219.77&        4286.93\\
   26&&        2934.69&        2986.58&        3036.24&        3082.83&&
               4230.79&        4305.15&        4375.88&        4443.28\\
   27&&        3044.15&        3096.07&        3145.94&        3192.80&&
               4386.60&        4461.07&        4532.20&        4600.01\\
   28&&        3153.68&        3205.77&        3255.68&        3302.66&&
               4542.70&        4617.33&        4688.59&        4756.82\\
   29&&        3263.18&        3315.36&        3365.50&        3412.70&&
               4698.84&        4773.76&        4845.36&        4913.76\\
   30&&        3372.84&        3425.10&        3475.26&        3522.63&&
               4855.34&        4930.30&        5002.17&        5070.99\\
   31&&        3482.42&        3534.76&        3585.10&        3632.58&&
               5011.94&        5087.14&        5159.10&        5228.14\\
   32&&        3592.07&        3644.40&        3694.70&        3742.31&&
               5168.62&        5243.95&        5316.22&        5385.43\\
   33&&               &               &               &               &&              
               5325.52&        5400.85&        5473.22&        5542.77\\              
\\ \hline
\end{tabular}
\end{table*}

\begin{figure}
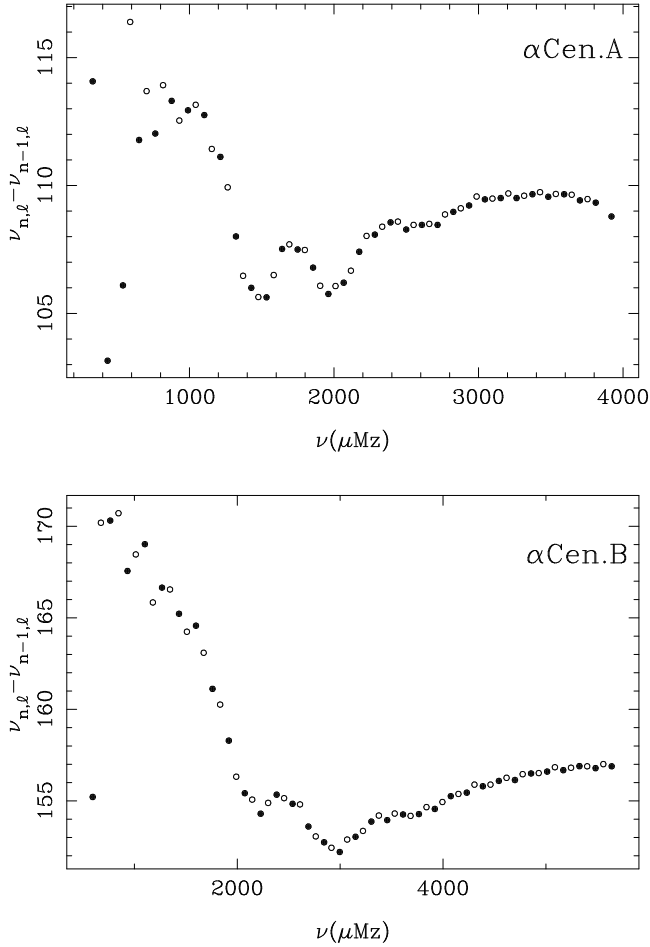

 \vbox{ \psfig{figure=10088.f1,height=6cm,angle=0}
\vspace{0.5truecm}
 \psfig{figure=10088.f2,height=6cm,angle=0}}
\caption{
Variations of the large frequency separations between modes of consecutive 
radial order
$\Delta\nu_{n,\ell} = \nu_{n,\ell}-\nu_{n-1,\ell}$
for $p$-modes of degree $\ell= 0$ (full point) and $\ell= 1$ (open point)
as a function of the 
frequency for $\alpha$\,Cen\,A \& B (models $\rm A_{BV}$ \& $\rm B_{BV}$).
The relation between the frequencies and the radial order are
taken from Table~\ref{tab:seisab}.
As asymptotically predicted, $\Delta\nu_{n,\ell}$ is almost constant 
at high frequency.}
\label{fig:f2}
\end{figure}

\begin{figure}
\psfig{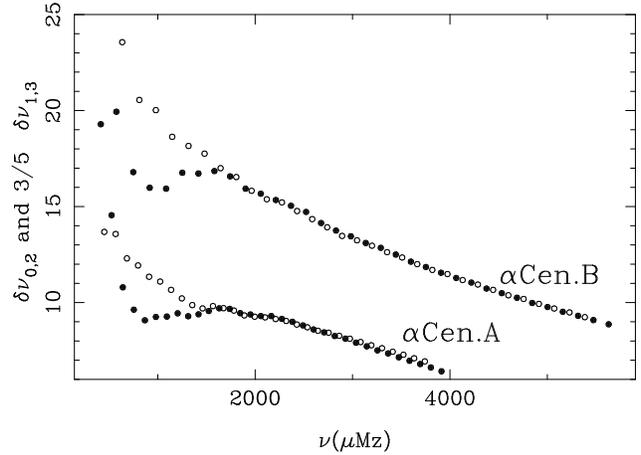}
\caption{
Variations of the small frequency differences 
$\delta\nu_{0,2}$ and 
$\frac 35\delta\nu_{1,3}$
 as a function of the 
frequency for $\alpha$\,Cen\,A \& B (models $\rm A_{BV}$ \& $\rm B_{BV}$). 
Same symbols as in Fig.~\ref{fig:f2}.}
\label{fig:f3}
\end{figure}

\begin{figure}
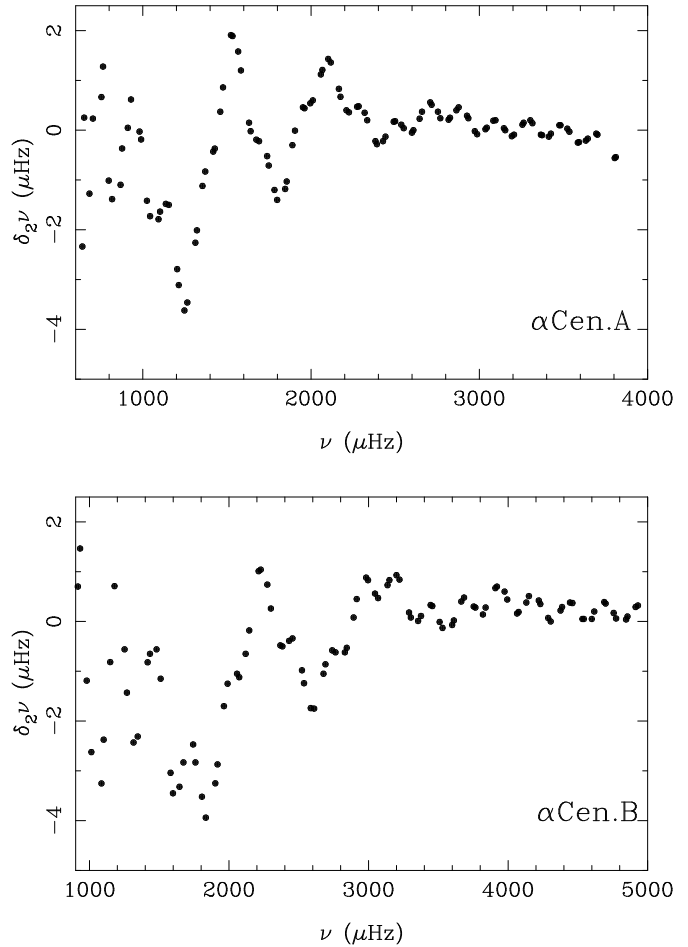

 \vbox{ \psfig{figure=10088.f4,height=6cm,angle=0}
\vspace{0.5truecm}
 \psfig{figure=10088.f5,height=6cm,angle=0}}
\caption{
Variations of the differences of frequencies
$\Delta_{n, \ell}=\nu_{n,\ell}-2\nu_{n+1,\ell}+\nu_{n+2,\ell}$
for $p$-modes of degree $\ell= 0, 1, 2, 3$ as a function of the
frequency for $\alpha$\,Cen\,A \& B (models $\rm A_{BV}$ \& $\rm B_{BV}$).
 The ``period'' $\cal P$
of the small oscillation is $203\,\mu$Hz for $\rm A_{BV}$ and $255\,\mu$Hz for
$\rm B_{BV}$ }
\label{fig:f4}
\end{figure}

\subsection{Seismological analysis of $\alpha$\,Cen\,A \& B}
The stars $\alpha$\,Cen\,A \& B are  solar-like stars.
The oscillations of such stars may be
stochastically excited by the convection as in the case of the Sun.
The amplitudes of solar-like oscillations have been estimated for stars
of different masses and ages (Houdek~\cite{h96}, Houdek et al.
\cite{h99}).
The models we have calibrated are close  in  the
H-R diagram, but have different internal structure  and they could be
discriminated with the help of seismology.
The properties of the stellar oscillations are related
to  the variation along the radius
of the sound speed $c$ and of the Brunt-V\"aiss\"al\"a frequency $N$
 (Christensen-Dalsgaard \& Berthomieu~\cite{cb91}).
The $p$-mode frequencies are mainly related to $c$, while the $g$-modes,
with lower frequencies, are determined essentially by $N$.
We have computed for all our models
the adiabatic  frequencies of  oscillation for modes of low degrees
$\ell$=0 to 3, which may be detected by  future observations.
Table~\ref{tab:seisab} presents the $p$-mode  frequencies for
the models $\rm A_{BV}$ and $\rm B_{BV}$. As in the solar case
(Provost et al.~\cite{pbm00}),
the lowest frequency $p$-mode oscillations of  $\alpha$\,Cen\,A
 have a mixed character between $p$- and $g$-mode behavior and are very 
sensitive to the structure of the central stellar regions.

\subsubsection{$p$-mode oscillations}\label{sec:pmode}
The frequencies  of $p$-modes of given degree $\ell$
are almost quasi-equidistant.
Figure~\ref{fig:f2} represents the variation of
$\Delta\nu_{n, \ell}=\nu_{n, \ell} - \nu_{n-1,\ell}$ as a function
of the frequency for
models $\rm A_{BV}$ and $\rm B_{BV}$; $n$ is the radial order of the
mode. $\Delta\nu_{n, \ell}$ represents the large spacing
between the mode frequencies. It is roughly constant at large frequencies,
larger than $2300\,\mu$\,Hz for $\alpha$\,Cen\,A and 
$3500\,\mu$\,Hz for $\alpha$\,Cen\,B.
Below these frequencies it varies within 10\%.
In the lower frequency range, $\Delta\nu_{n, \ell}$
 has an oscillatory
behavior which is the signature of the helium ionization zone (e.g.
Gough~\cite{g91}).
We note that it is important to take into account such a variation
while searching  peak  equidistancy  in the $p$-mode power spectrum.

Another characteristic quantity of the oscillation spectrum is the
small separation, i.e. the difference between the frequencies of modes
with a degree of same parity and with consecutive
radial order:
\[\delta\nu_{\ell, \ell+2} = \nu_{n+1,\ell} - \nu_{n,\ell+2}.\]
This small quantity is very sensitive to the structure of the core
mainly to its hydrogen content.
Asymptotic analysis predicts that it is proportional to $\ell(\ell+1)$.
Figure \ref{fig:f3} gives $\delta\nu_{0, 2}$ and $\frac35\times\delta\nu_{1,3}$
 as a function of the frequencies for the
models
$\rm A_{BV}$ and $\rm B_{BV}$. For a given model, these two quantities
are very close at
high frequencies, as expected from asymptotic approximation and they
vary
almost linearly with  frequency for radial orders $n$ larger than about
16 for $\alpha$\,Cen\,A and 10 for $\alpha$\,Cen\,B.

In the high
frequency range the large and small frequency spacings which characterize
the $p$-mode spectrum are usually estimated by analytical fits 
of the frequencies and of the small frequency separations.
The numerical frequencies are fitted by the following polynomial
expression (Berthomieu et al.~\cite{b93}):
\[ \nu_{n, \ell} =  \nu_{0\ell} + \Delta\nu_\ell (n+{\ell\over 2}-n_0)
+a_\ell  (n+{\ell\over 2}-n_0)^2.\]
The quantity $\delta\nu_{\ell,\ell+2}$ varies almost
linearly with the frequency or the radial order. The small spacing
is  estimated  by using the  fit:
\[\delta\nu_{\ell, \ell+2} \sim
\overline{\delta\nu}_{\ell, \ell+2}+S_\ell (n-n_0).\]
As in  the solar case,  we consider a set of 9 modes centered at
$n_0 \sim$ 21, which corresponds approximately
to the middle of the range of the
expected excited frequencies (i.e. $1 - 3$\,mHz) for $\alpha$\,Cen\,A,
according to Houdek~(\cite{h96}).
The fitted coefficients $a_\ell$ are very small.
The quantity $\Delta\nu_\ell$ does not much depend on the degree,
so that $\Delta\nu_\ell\sim \Delta\nu_0$
and it characterizes the large frequency spacing.
The small spacings are conventionally measured by
$\overline{\delta\nu}_{\ell, \ell+2}$.
Table~\ref{tab:seis} shows the quantities $\nu_{00}$, $\Delta\nu_0$,
$\overline{\delta\nu}_{0,2}$ and
$\overline{\delta\nu}_{1,3}$ which have been computed for the models of
$\alpha$\,Cen\,A \& B
given in Table~\ref{tab:mod}.

These large and small separations of frequency, which characterize the
$p$-mode oscillation spectrum, depend on the stellar mass and age and
slightly decrease with the age (Christensen-Dalsgaard~\cite{c84},
Audard et al.~\cite{apc95}). $\nu_{00}$ and $\Delta\nu_0$
are mainly related to the envelope structure of the stellar model and
they vary proportionally to $M^{\frac12} R^{-\frac32}$,
while the values of $\overline{\delta\nu}_{0,2}$ and
$\overline{\delta\nu}_{1,3}$ reflect the
 structure in the core.

The computed values of $\Delta\nu_{0}$ are respectively
smaller ($respt.$ larger) than
the solar ones due to larger ($respt.$ smaller) masses of
$\alpha$\,Cen\,A ($respt.$ $\alpha$\,Cen\,B).
These values do not depend much on the description of the convection or on
the convective core overshoot, as can be seen from a comparison of the models
$\rm A_{BV}$, $\rm A_{CM}$, $\rm A_{ov}$ and  $\rm B_{BV}$, $\rm
B_{CM}$, $\rm B_{ov}$.
On the contrary, the value of  $\Delta\nu_{0}$ obtained for the oldest
model  $\rm A_{GD}$
of $\alpha$\,Cen\,A, which has also a larger mass, is significantly smaller
by about $5\,\mu$Hz. The result is opposite for  $\alpha$\,Cen\,B.
All these differences can be accounted for by the differences in mass and
radius.

The small separations  $\overline{\delta\nu}_{0,2}$ and
$\overline{\delta\nu}_{1,3}$ decrease with
stellar age and mass (Christensen-Dalsgaard~\cite{c84}).
The  values of  $\overline{\delta\nu}_{0,2}$ and
$\overline{\delta\nu}_{1,3}$
obtained for our calibrated models are
comparable to those  of Pourbaix et al.~(\cite{pnn99}).
As expected, they decrease with the age 
for the models without core overshoot, and thus are
significantly smaller for the $\rm A_{GD}$  model of $\alpha$\,Cen\,A.
Future asteroseismic observations
could help to discriminate between such models.

The lower panel of Table~\ref{tab:seis} presents
the variations $\Delta[\Delta\nu_0]$,
$\Delta[\overline{\delta\nu}_{0,2}]$ and
$\Delta[\overline{\delta\nu}_{1,3}]$  for the models computed with
the $\rm MLT_{BV}$ theory and extreme modeling parameters
$\alpha$, $t_{\rm\alpha\,Cen}$, $Y_{\rm i}$ and $\rm[\frac{Fe}H]_i$ within
the confidence domains
presented in Table~\ref{tab:glob}. The variations of $\Delta\nu_0$ with
$\alpha$,
$t_{\rm\alpha\,Cen}$ and $Y_{\rm i}$ are essentially due to the difference
of model radii.
The small spacings $\overline{\delta\nu}_{0,2}$ and
$\overline{\delta\nu}_{1,3}$ depend
mainly on the age and are less sensitive to the differences in radius.

According to Gough~(\cite{g91}), further information on the stellar
structure can be provided by low degree oscillations, from the
second order difference of frequencies:
\[\Delta_{n,\ell}=\nu_{n,\ell}-2\nu_{n+1,\ell}+\nu_{n+2,\ell}.\]
When plotted as a function of the frequency (Figure~\ref{fig:f4}),
this quantity has a sinusoidal behavior with two different ``periods''
of order $210\,\mu$Hz and $700\,\mu$Hz. The larger period has the
largest
amplitude and is due to the rapid
variation of the adiabatic index $\Gamma_1$ in the HeII ionization
zone.
Its contribution to the frequency is also clearly visible in
Fig.~\ref{fig:f2}.
The lower period ${\cal P}$ is an
indication for a discontinuity in the derivative of the sound velocity
at the base of the convection zone and is the inverse of twice the
travel time
of the sound from the surface to the base of the convection zone
(e.g. Audard \& Provost~\cite{ap94}):
\[{\cal P}^{-1} =2 \int_{R_{\rm cz}}^{R_\star}\frac{dr}c.\]
The predicted value of ${\cal P}$
is larger for $\alpha$\,Cen\,B than for $\alpha$\,Cen\,A,
mainly due to a deeper convection zone for $\alpha$\,Cen\,B (see
Table~\ref{tab:mod}).
The corresponding amplitude is sensitive
to different processes, like convective penetration below
the convection zone (e.g. Berthomieu et al.~\cite{bmpz93}).

\subsubsection{$g$-mode oscillations}
In the low frequency range, the frequencies are mainly determined by the
 Brunt-V\"aiss\"al\"a frequency $N$. The period of low degree gravity
modes is proportional to a characteristic period:
\[\nu_{n, \ell} \sim P_0^{-1} (n+\varepsilon_\ell)=
\frac1{2\pi}\int_{\rm R_{co}}^{\rm R_{zc}}\frac Nr\,dr.\]
The integral, which defines $P_0$, is taken in the inner radiative zone,
i.e. from the radius of the convective core ${\rm R_{co}}$ for $\alpha$\,Cen\,A,
or from the center for $\alpha$\,Cen\,B,
 to  the base of the external convection zone ${\rm R_{zc}}$.
For stars without convective core, like the Sun or $\alpha$\,Cen\,B,
$\varepsilon_\ell$ depends on the degree of the oscillation,
 $\varepsilon_\ell=\ell/2+\overline\varepsilon$;
for stars with convective core (like  $\alpha$\,Cen\,A)
$\varepsilon_\ell=\overline\varepsilon$ (Christensen-Dalsgaard~\cite{c84}).
Table~\ref{tab:seis} gives the values of $P_0$ for the models of
$\alpha$\,Cen\,A \& B.
When the calibration of $\alpha$,Cen\,A is made with a core overshoot
(which increases the extent of the convective core) $P_0$
is  significantly larger (by 23\%).
The other differences between the values of $P_0$ for the
models of $\alpha$\,Cen\,A \& B are 
accounted for by the differences in mass and in radius.

\section{Discussion and conclusion}\label{sec:dis}
Detailed evolutionary calculations of the visual binary $\alpha$\,Centauri,
including the pre main-sequence
have been performed using the recent mass determinations of Pourbaix et
 al.~(\cite{pnn99}). Models have been constructed using the CEFF equation of
 state, OPAL opacities, NACRE thermonuclear reaction rates and
 microscopic diffusion.
We have revisited the effective temperatures, surface gravities and 
metallicities, using published spectroscopic data and
taken these quantities as observational constraints. 
We have determined the most reliable solution 
within the confidence domains of the
observable constraints via a $\chi^2$ minimization.
Each solution is characterized by
$\wp=\{t_{\rm\alpha\,Cen},Y_{\rm i},[{\rm\frac{Fe}H]_i},\alpha_{\rm A},\alpha_{\rm B}\}$,
where $t_{\rm\alpha\,Cen}$ is the
 age of the system, $Y_{\rm i}$ the initial helium content,
 $[{\rm\frac{Fe}H]_i}$ the initial metallicity and 
 $\alpha_{\rm A}$ and $\alpha_{\rm B}$ the convection parameters
 of each star model.
 We obtained calibrations using different convection theories
and adapted values for the mixing-length parameter of each component.

With the basic B\"ohm-Vitense~(\cite{b58}) mixing-length
 theory of convection,
 we derived $\wp_{\rm BV}=\{\rm 2710\,Myr, 0.284,0.257,1.53,1.57\}$.
With a convective core overshoot of $0.20\,H_{\rm p}$,
  we obtained $\wp_{\rm ov}=\{\rm 3530\,Myr, 0.279,0.264,1.64,1.66\}$.
With the Canuto \& Mazitelli~(\cite{cm91},~\cite{cm92}) convection
theory, we get  
 $\wp_{\rm CM}=\{\rm 4086\,Myr, 0.271,0.264,0.964,0.986\}$.
 Using the mass values and
 the observational constraints of the ``best'' model of
 Guenther \& Demarque~(\cite{gd00}),
 with the basic mixing-length theory,  
  we obtained $\wp_{\rm GD}=\{\rm 5638\,Myr, 0.300,0.296,1.86,1.97\}$.

We have also performed $\chi^2$ minimizations forcing the use of
a unique value
for the convection parameters of models of $\alpha$\,Cen\,A \& B.
We have not reported the results
as they are not significantly different from the former ones.

\medskip

The most striking fact in our results is the small values obtained
for the ages
which are noticeably smaller than previous studies, except
those of Boesgaard \& Hagen~(\cite{bh74}) and
Pourbaix et al.~(\cite{pnn99}). With respect to the recent
models of Guenther \& Demarque~(\cite{gd00})
this is due mainly to the mass discrepancies
resulting from the small differences in distances.
The $\sim5.5\%$ increase between the masses of Pourbaix et al.~(\cite{pnn99})
we used and those used by Guenther \& Demarque~(\cite{gd00}) results from
by the 1.5\% disparity between the revised Hipparcos parallax
(S\"oderhjelm~\cite{s00}) used by Guenther \& Demarque
and the orbital parallax determination of Pourbaix et al.
As revealed by the $\chi^2$ minimization, in the case
of the present calibration
of $\alpha$\,Cen, the determination of the age
appears to be more sensitive to the mass differences than to the
basic observational atmospheric constraints namely, effective temperature,
gravity -- or luminosity -- and metallicity. According to our results,
 no really satisfactory criterion 
allows to discriminate among different sets of modeling parameters:
they all generate models which fit the observational constraints
based on H-R diagram analysis and metallicities.
A naive argument, however, can plead in favor of a subsolar age: as the
$\alpha$\,Cen binary system is formed of metal enriched material,
it formed from more processed interstellar matter than the Sun;
this may indicate that $\alpha$\,Cen is younger than the solar system.

\medskip

For the models computed with the basic
$\rm MLT_{BV}$ theory and fitted on our observational constraints,
the mixing-length parameters
are noticeably smaller than the convection parameter $\alpha_\odot$ of the 
solar model calibrated with the same physics;
the differences are larger than expected from
the Ludwig et al.~(\cite{lfs99}) simulations.
For the models fitted to the observational constraints of the ``best'' model of
Guenther \& Demarque~(\cite{gd00}),
the values derived for $\alpha$ bracket the calibrated solar value.
For all models computed with $\rm MLT_{BV}$ theory we note that
the smaller the age, the smaller the mixing-length parameters.

For models computed with $\rm MLT_{CM}$  convection theory,
in accordance with Canuto \& Mazitelli~(\cite{cm91},~\cite{cm92}) we obtain
mixing length parameters both close to unity and to the convection parameter
$\alpha_\odot$ of the solar model calibrated with the same physics,
$\alpha_{\rm A}\simeq\alpha_{\rm B}\simeq\alpha_\odot\simeq 1$.
This indicates that the convection theory of 
Canuto \& Mazitelli may
support the assumption of a universal convection parameter
and it seems to provide better fits to observations,
as already pointed out by helioseismology as aforesaid in Sect.~\ref{sec:mlt}.

\medskip

For the $\alpha$\,Cen\,A models fitted on our observational constraints,
as in the solar model,
microscopic diffusion alone is not efficient enough to account for
the observed surface lithium depletion.
This may indicate that an unknown physical process, at work
beneath the outer convection zone, reinforces the microscopic diffusion
and gravitational settling
to transport the material down to the lithium burning zone.
For $\alpha$\,Cen\,B, the only available observation of
the surface lithium abundance is an upper limit
(Chmielewski et al.~\cite{cfcb92}): all $\alpha$\,Cen\,B
models predict values compatible
with this observation except model $\rm B_{BV}$.
The surface lithium depletion of models calibrated with the
Guenther \& Demarque~(\cite{gd00}) observational
 constraints are close to the observations and, in
that sense, they appear to be more satisfactory than models
fitted on our observational constraints: in these models
there is no need for additional physical process
 to transport the lithium to its burning zone. But this imply that the small 
lithium depletion found in the standard solar model and the too large efficiency
of the microscopic diffusion in more massive
stars result from physical processes which are not active in
$\alpha$\,Cen, though masses and rotation status are close to solar ones.

All our models with large masses have initial helium
contents close to the solar value
${Y_\odot-0.003}\leq Y_{\rm i}\leq Y_\odot+0.010$, while
the models with low masses have a high initial helium
content $Y_{\rm i}=Y_\odot+0.026$. With a primordial
helium abundance of $Y_0\approx 0.235$ we get a galactic enrichment of
$\frac{\Delta Y}{\Delta Z}\approx2.0$, both for the Sun and low mass
models and
$\frac{\Delta Y}{\Delta Z} =1.2-1.6$ for high mass
models. The differences in $\frac{\Delta Y}{\Delta Z}$ between our
$\alpha$\,Cen models and the Sun are compatible with the scattering
found in the solar
neighborhood (Pagel \& Portinari~\cite{pp98}) and in other binary system
calibrations (Fernandes et al.~\cite{flbm98}).

We have computed the large and small frequency spacings of acoustic
oscillations for all
the models. The large separation $\Delta\nu_0$ for the three models
with large masses are within $2\,\mu$Hz,
despite their differences in age and physics. They are such that
$\Delta\nu_0\sim 107-108\,\mu$Hz. The variations of $\Delta\nu_0$
taking into account the uncertainties in the observable constraints,
effective temperatures, gravities and metallicities, are of about
$\pm 1\,\mu$Hz. For the model $\rm A_{GD}$ of small mass, the large separation,
$\Delta\nu_0\sim 102\,\mu$Hz, is in agreement with the value of
Guenther \& Demarque~(\cite{gd00}) and significantly lower
for our other models.
 The differences in  $\Delta\nu_0$ are mainly explained 
 by the differences in mass and radius.

For the three models with large masses
the small separations $\overline{\delta\nu}_{0,2}$ vary from 7.5 to $9.1\,\mu$Hz
for $\alpha$\,Cen\,A and are of about $12\,\mu$Hz for $\alpha$\,Cen\,B.
 For each model the variations in
$\overline{\delta\nu}_{0,2}$ within the observable uncertainties are
$\pm 1\,\mu$Hz. For  the model $\rm A_{GD}$, $\overline{\delta\nu}_{0,2}$
is much smaller and in agreement with the value of
Guenther \& Demarque.
All these differences in $\overline{\delta\nu}_{0,2}$ are mainly related
 to the differences in central hydrogen content, hence in the age.

These results show that the determination of $\Delta\nu_0$ and
$\overline{\delta\nu}_{0,2}$ by seismological observations would
help to discriminate between the models of $\alpha$\,Cen\,A computed with
different masses and to confirm or not the new determination of the masses
by Pourbaix et al. This implies an improvement
of the accuracy of the
observables used to constrain the calibrated models of $\alpha$\,Cen\,A
\& B. Concerning the comparison with the seismic observations,
the large splitting estimated by Pottasch et al.~(\cite{pbh92})
and Edmonds \& Cram~(\cite{ec95}) favor Pourbaix et al~(\cite{pnn99}).
masses.
The different possible estimations for the large and small spacings
$\Delta\nu_0$ and $\overline{\delta\nu}_{0,2}$ by Kjeldsen
et al.~(\cite{kbfd99}) do not allow to discriminate between the
models. We note, however, that their estimation
$\Delta\nu_0=100.8\,\mu$Hz and $\overline{\delta\nu}_{0,2}=11.7\,\mu$Hz is
highly improbable.

\medskip

We conclude that, even for $\alpha$\,Cen, the best known binary system,
the models are not strongly enough constrained by the available
astrometric, photometric and spectroscopic data. 
In order to deeply test stellar physics additional
information on the internal structure is needed.
Up to now ground-based observations give tentative evidence
for acoustic oscillations in $\alpha$Cen A.
In a few years from now, one can expect that asteroseismology from space
e.g. COROT (Baglin et al.~\cite{bc98}), MONS (Kjeldsen et al.~\cite{kbcd99}) 
and MOST (Mattews~\cite{m98}) missions
and from ground, e.g.
Concordiastro (Fossat et al.~\cite{fgv00}), will provide data accurate enough
to improve our knowledge of stellar interiors.

\begin{acknowledgements}
We would like to express our thanks to the referee, Dr. Pourbaix,
for helpful advices.  
This research has made use of the Simbad data base, operated at
CDS, Strasbourg, France.
This work has been performed using the computing facilities 
provided by the OCA program
``Simulations Interactives et Visualisation en Astronomie et M\'ecanique 
(SIVAM)''.
\end{acknowledgements}

\end{document}